\documentclass[
  aps,
  prd,
  reprint,
  superscriptaddress,
  nofootinbib,
  floatfix
]{revtex4-2}

\usepackage{amsmath}
\usepackage{graphicx}
\usepackage{xcolor}
\usepackage{url}
\usepackage{comment}

\graphicspath{{./images/}}

\begin{document}

\title{GW-YOLO: Multi-transient Segmentation in LIGO Using Computer Vision}

\author{Siddharth Soni}
\email{siddhars@ucr.edu}
\affiliation{LIGO Laboratory, Massachusetts Institute of Technology, Cambridge, Massachusetts 02139, USA}
\altaffiliation{Present address: Department of Physics and Astronomy, University of California, Riverside, California 92521, USA}

\author{Nikhil Mukund}
\email{nikhil@iucaa.in}
\affiliation{LIGO Laboratory, Massachusetts Institute of Technology, Cambridge, Massachusetts 02139, USA}
\altaffiliation{Inter-University Centre for Astronomy and Astrophysics, Pune 411007, Maharashtra, India}

\author{Erik Katsavounidis}
\email{kats@mit.edu}
\affiliation{LIGO Laboratory, Massachusetts Institute of Technology, Cambridge, Massachusetts 02139, USA}

\begin{abstract}
Time series data and their time-frequency representation from gravitational-wave interferometers present multiple opportunities for the use of artificial intelligence methods associated with signal and image processing.
Closely connected with this is the real-time aspect associated with gravitational-wave interferometers and the astrophysical observations they perform; the discovery potential of these instruments can be significantly enhanced when data processing can be achieved in O(1s) timescales.
In this work, we introduce a novel signal and noise identification tool based on the YOLO (You Only Look Once) object detection framework. For its application into gravitational waves, we will refer to it as GW-YOLO. This tool can provide scene identification capabilities and essential information regarding whether an observed transient is any combination of noise and signal. Additionally, it supplies detailed time-frequency coordinates of the detected objects in the form of pixel masks, an essential property that can be used to understand and characterize astrophysical sources, as well as instrumental noise.
The simultaneous identification of noise and signal, combined with precise pixel-level localization, represents a significant advancement in gravitational-wave data analysis. 
Our approach achieves a 53\% detection efficiency for binary black hole
signals in the signal-to-noise ratio (SNR) 12--15 range when such signals overlap with transient
noise artifacts, increasing to more than 75\% in the SNR 15--18 range.
For binary neutron star signals overlapping with transient noise artifacts,
the detection efficiency reaches 57\% in the SNR 30--33 range and increases
to 87\% in the SNR 39--42 range.
To our knowledge, this is the first quantitative assessment of the ability
to detect astrophysical signals overlapping with realistic instrumental
noise present in gravitational-wave interferometers.
It also presents a fully automated, low-latency pipeline for producing
pixel-level segmentation masks for individual noise transients and
astrophysical signals, enabling further automation of event validation
and downstream noise-mitigation procedures.
\end{abstract}

\maketitle

\section{Introduction}\label{sec:level1}

Gravitational waves (GWs) have ushered a new era in observing and learning about our universe. Following the breakthrough first direct detection of the merger of a pair of black holes in 2015 \citep{LIGOScientific:2016aoc} by the Laser Interferometer Gravitational-wave Observatory (LIGO) \citep{LIGOScientific:2014pky}, hundreds of detections of astrophysical sources have been made by LIGO and the international network of gravitational-wave detectors that also includes Virgo \citep{VIRGO:2014yos} and KAGRA \citep{kagra}.
Gravitational-wave data from these instruments correspond to time series sampled at frequencies from few Hz to few kHz. The primary astrophysical
data product is the calibrated gravitational-wave strain channel, which
reconstructs the differential arm-length response of the interferometer and
therefore contains the expected signatures of astrophysical gravitational-wave
signals with the highest sensitivity
\citep{aLIGO:2020wna, Capote:2024rmo}. Additionally the instruments record a wealth of environmental and instrumental sensors meant to capture environmental disturbances as well as the overall state of the interferometry. These auxiliary channels provide essential information for assessing the quality of data (and validity of detected signals as astrophysical) as well as for controlling the interferometry \citep{LIGO:2026ika, LIGO:2024kkz, LIGO:2021ppb, Aso:2013eba}.

Machine learning has emerged as a transformative tool for extracting structure, patterns, and information from high-dimensional data, particularly in the domain of image processing. Techniques such as deep convolutional neural networks have shown remarkable success in tasks like object detection, segmentation, and classification across a variety of scientific and engineering applications. In gravitational-wave astronomy, where detector data is often visualized as time-frequency images (e.g., spectrograms or Q-transforms \citep{chatterji2004_qtransform}), machine learning offers a powerful framework for identifying and characterizing transient phenomena that may be difficult to isolate through traditional signal processing methods~\citep{biswas2013application,Zevin2017GravitySpy,Mukund2017Transient,powell2017classification,George2018DeepFiltering,Gabbard2018Matching,Razzano2018ImageBased,Coughlin2019Discovering}.
The time-frequency representation of data is an essential step in analyzing gravitational-wave data as both instrument noise and astrophysical sources manifest themselves with distinct frequency content and overall patterns that capture the mechanism of noise generation and interference with the instruments and the astrophysical process that emits gravitational waves, respectively.

LIGO and its partner observatories produce vast amounts of complex data. The primary astrophysical sources identified so far are from mergers of black hole and neutron star pairs. Beyond these astrophysical signals, gravitational-wave instruments record a wide array of non-Gaussian, non-stationary noise transients \citep{LIGOScientific:2014pky, VIRGO:2014yos, KAGRA:2021vkt, LIGOScientific:2019hgc}. These non-astrophysical transients, often referred to as glitches, can mimic or obscure real gravitational-wave signals, complicating both detection and (astrophysical) interpretation \citep{LIGOScientific:2016gtq}. While traditional pipelines rely on model-driven matched filtering~\citep{sathyaprakash1991,allen2012}, recent advances in machine learning provide an opportunity to approach the problem from a data-driven, morphology-aware perspective \citep{Benoit2025Aframe}. In particular, the use of deep learning for image-based classification and segmentation holds promise for disentangling signal and noise, localizing features in time-frequency space, and enabling real-time inference pipelines. The work we report here builds upon these developments, applying modern machine learning methods to improve transient identification and signal discrimination in gravitational-wave data.

In order to process gravitational-wave data, several sophisticated computational techniques have been developed, including matched filtering algorithms, machine learning-based classifiers, and transient noise mitigation methods \citep{Davis:2018yrz, Cornish:2014kda, Zevin2017GravitySpy}. These techniques help improve detection sensitivity and event characterization, enabling improved astrophysical interpretations of observed signals.

\begin{figure}
    \centering
    \includegraphics[width=\linewidth]{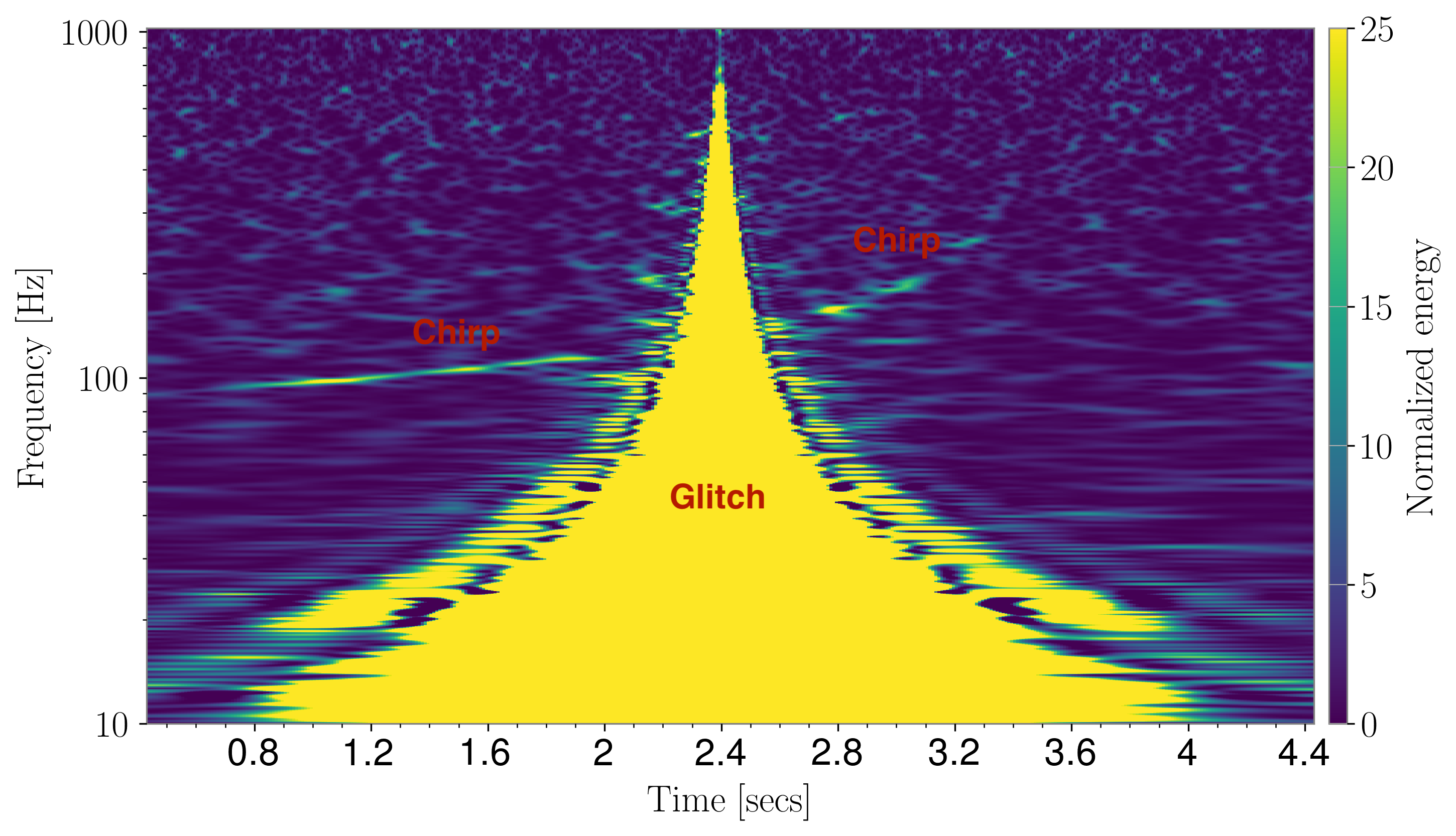}
    \caption{Time-frequency spectrogram of the binary neutron star event GW170817~\citep{abbott2017gw170817}. The astrophysical event corresponds to the time-frequency track sloping up in frequency as time elapses (and labeled as ``chirp''). A loud glitch at the time of this event that adversely impacted the quality of data at the LIGO-Livingston detector can also be seen dominating the low frequency (and beyond) response of the instrument (and labeled as ``glitch''). Such overlaps or even in near proximity of noise and signal can challenge significantly the detection and parameter estimation of astrophysical events.}
    \label{fig:bns_glitch_event}
\end{figure}

GW170817, the first ever binary neutron star merger event which electromagnetic counterpart was observed using multiple telescopes~\citep{abbott2017gw170817}, had a significant non-astrophysical overlapping glitch in the LIGO Livingston observatory as shown in Fig. \ref{fig:bns_glitch_event}. This glitch had to be carefully and manually removed from the gravitational-wave strain data using an analysis framework called BayesWave~\citep{Cornish:2014kda} prior to undertaking the event reanalysis including its parameter estimation studies. With the expected improvement in the detector sensitivity in the coming years, we expect to see more such nearby occurrences of binary merger events themselves and or alongside glitches. 
Such time-frequency overlaps present the prototypical case for the need of multi-object detection in gravitational-wave data, we well as for full automation and real-time application.

The improved low-frequency sensitivity of planned third-generation (3G) detectors, such as the Cosmic Explorer (CE)~\citep{evans2021horizon,PhysRevD.91.082001} and the Einstein Telescope (ET)~\citep{punturo2010einstein}, is expected to dramatically increase the number of detected events. These detectors will be capable of capturing hundreds of binary neutron star (BNS) events each day and several binary black hole (BBH) events every hour~\citep{ Himemoto2021Impacts}. However, the overlap of low signal-to-noise ratio (SNR) events (with SNR $\leq$ 4) near stronger events may introduce biases in the parameter estimation processes, as the overlap between the GW signal and glitch in time and frequency can affect the results. Additionally, gravitational microlensing caused by stellar-mass compact objects can lead to delays of milliseconds to seconds for secondary signals, potentially resulting in interference from beating waveforms~\citep{nakamura1998gravitational,nakamura1999wave,takahashi2003wave}.

Previous studies investigating the impact of contamination from non-astrophysical transients on black hole and supernova signals have found that such interference can significantly impair parameter estimation and model selection~\citep{powell2018parameter}.
Other studies on overlapping gravitational-wave signals on parameter estimation, using both Fisher information matrix methods~\citep{Himemoto2021Impacts} and Bayesian inference techniques~\citep{Samajdar2021}, have shown that existing approaches are generally robust, except in rare cases involving nearly coincident coalescence times or large differences in signal-to-noise ratios. However, a comprehensive understanding of the effects of improved detector sensitivity at frequencies below 5 Hz, which lead to longer signal durations, and the influence of binary spin remains an open challenge.

In this paper, we study image segmentation of gravitational-wave time-frequency spectrograms and introduce a novel signal-versus-noise and multi-scene identification tool based on the YOLO (You Only Look Once) object detection framework~\citep{Redmon2016YOLO}. There have been recent efforts that focuses on a time-frequency rectangular region also known as ``bounding box" approach to discern between various non-astrophysical transients seen in the GW detector using the older YOLOv3 architecture~\citep{LIGO:P2500369}. Our work addresses the problem of simultaneous identification of transient noise and GW signals in the data using the more modern YOLOv11 architecture~\citep{ultralytics_yolo}.
This tool provides instance segmentation capabilities and essential information regarding whether an observed transient is noise, signal or a combination of them. By instance segmentation, we mean that the network not only detects whether a transient belongs to a given class (noise, signal, or both) but also produces a separate pixel-level mask for each individual occurrence (instance) of that class. 
The simultaneous identification of noise and signal, combined with precise pixel-level localization, represents a significant advancement in gravitational-wave data analysis. By rigorously validating our approach on real LIGO detector data, we demonstrate the potential of this method to improve the accuracy and reliability of gravitational-wave detections.

\section{\label{sec:level2}Transient noise}

\begin{figure*}[t]
  \centering
  \includegraphics[width=0.9\linewidth]{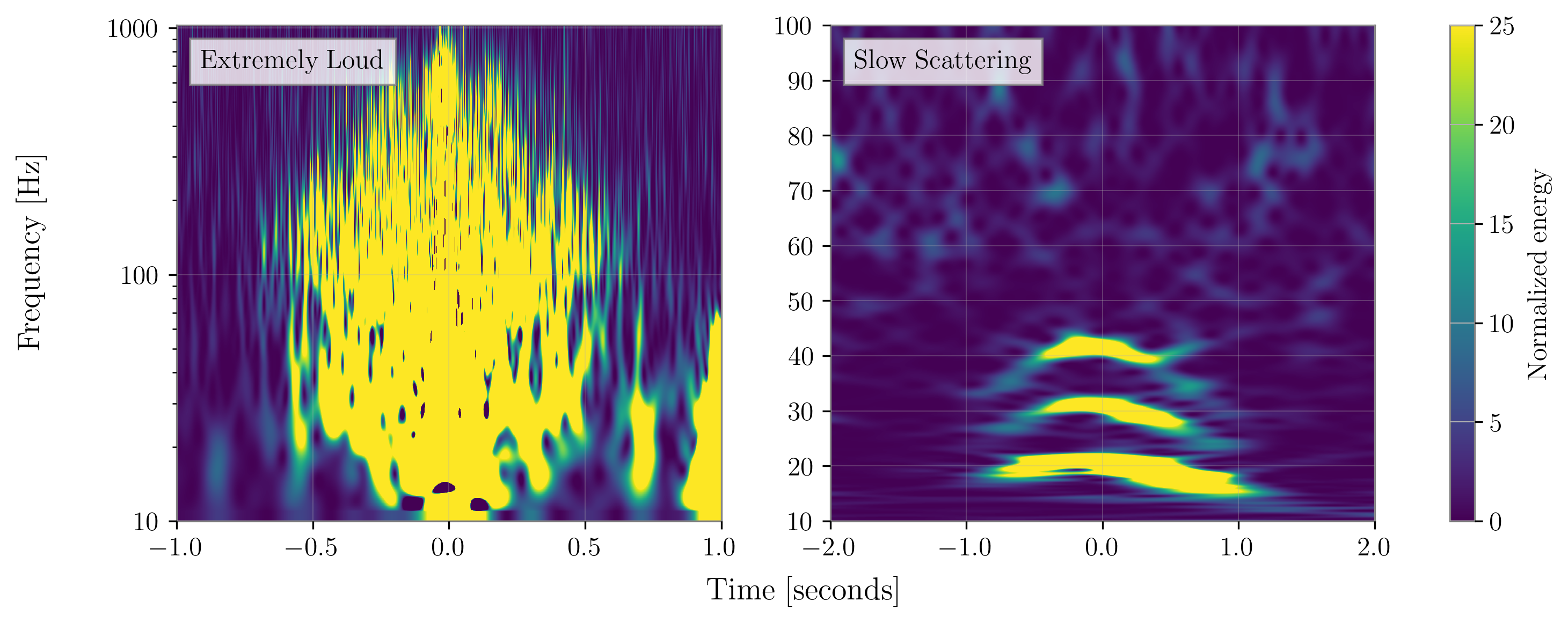}
  \caption{Time frequency spectrograms of two sample glitch categories ``extremely loud"  on the \emph{left} and ``scattered light" on the \emph{right} as identified by the GravitySpy algorithm  \citep{Zevin2017GravitySpy}. These are two of multiple transient noise classes that show up in the LIGO detector data.}
  \label{fig:glitch_scatter_loud}
\end{figure*}

Transient noise events in gravitational-wave detectors, commonly referred to as glitches, are brief (O(1s)) noise disturbances of non-Gaussian nature that adversely impact the data quality and the ability to observe gravitational waves \citep{LIGO:2021ppb, LIGO:2024kkz, Soni:2023kqq, Cabero:2019orq}. Fig \ref{fig:glitch_scatter_loud} shows two different classes of GravitySpy \citep{Zevin2017GravitySpy, Zevin:2023rmt, Wu:2024tpr} glitches that commonly show in LIGO data \citep{Glanzer_2023}.
These glitches can mimic or obscure genuine gravitational-wave signals, complicating their detection and accurate characterization. The presence of such noise instances increases the false alarm rate, potentially leading to misinterpretation or rejection of real astrophysical signals \citep{Macas:2022afm, Hourihane:2022doe}. Transients when present very close to the gravitational-wave signals, can complicate the parameter estimation analysis of signals and reduce our confidence in their astrophysical origin.  As mentioned earlier, Fig. \ref{fig:bns_glitch_event} shows GW170817, an astrophysical event where transient noise present very close to the signal complicated its detection and analysis \citep{abbott2017gw170817}. Effectively identifying and mitigating transient noise is thus crucial for enhancing the sensitivity and reliability of gravitational-wave detections, ultimately enabling more precise astrophysical insights and discoveries.

The LIGO, Virgo and KAGRA interferometers participated in the fourth Obseving run which began in May 2023 and ended in November 2025 \citep{Capote:2024rmo}. During the first part of O4, glitches in the LIGO instruments showed up at a rate of approximately 1 per min~\citep{LIGO:2024kkz}.
One of the primary tool to identify glitches in the data is an event trigger generator called Omicron~\citep{robinet2020_omicron}. It does so by identifying excess signal power in the data and annotates these instances with useful characteristics such as time of occurrence, signal-to-noise ratio, frequency, bandwidth (Q-factor), duration, amplitude etc. Depending on these characteristics, different glitches will have very different morphology when plotted in the time frequency spectrograms. GravitySpy which is a convolutional neural network-based tool, then separates these Omicron-identified transients into different categories based on how they appear in the time-frequency spectrograms \citep{Zevin2017GravitySpy, Soni:2021cjy, Glanzer_2023}.

\subsection{Q-transform}
Throughout this paper, we use Q-transforms of transient events (both noise and astrophysical signals) to create spectrograms of our data. These spectrograms also form the basis of our training dataset. 
The Q-transform is a time-frequency analysis technique widely used in gravitational-wave data analysis to visualize transient features in detector strain data~\citep{chatterji2004_qtransform, LIGO:2021ppb, gwpy_3_1_0}. Figures \ref{fig:bns_glitch_event} and \ref{fig:glitch_scatter_loud} are examples of Q-transform. It projects the time-series data x(t) onto a basis of sine-Gaussian wavelets, providing a high-resolution spectrogram that preserves both temporal and frequency localization. The multi-Q transform consists of multiple time-frequency planes or spectrograms, and each plane has time-frequency tiles with constant ratio of frequency to bandwidth (Q).  
The algorithm selects the plane containing the maximum-energy tile, thereby optimizing the representation for transient events.  This optimized constant Q spectrogram is then interpolated in order to provide a high resolution image of the transient features in the data. Notably, Q optimization is biased toward maximizing energy, which may not always be ideal for detecting low-SNR astrophysical signals corresponding to binary compact mergers in the presence of glitches. 

The Q-transform coefficients X are obtained from:
\begin{equation}
X(\tau,\phi,Q)
= \int_{-\infty}^{\infty}
  x(t)\,
  w\bigl(t-\tau,\phi,Q\bigr)\,
  \mathrm e^{-2\pi i \phi t}\,
  \mathrm d t
\end{equation}

and represent the average signal amplitude in the time-frequency tile. The signal energy is computed from these coefficients.  The tile has central time value of $\tau$, and central frequency $\phi$ and Q is the ratio of central frequency to the frequency bandwidth.

\section{YOLO}
YOLO (You Only Look Once) is a powerful and efficient deep learning framework initially designed for object detection tasks. It is widely recognized for its real-time performance and accuracy~\citep{Redmon2016YOLO}. It is being used for many different tasks including crop/weed detection, water leak recognition in tunnels and for real-time object detection in autonomous driving \citep{crop_weed_yolo, water_leak_yolo, yolo_obj_detection}. YOLO has also been used in LIGO for the detection of point absorber in test mirrors for gravitational-wave interferometers \citep{Goode:2024ccn}. 
By employing a single neural network to simultaneously predict rectangular regions (bounding boxes that tightly enclose potential objects) and the class probabilities for those regions across the entire input image, YOLO significantly streamlines the detection process.Recently, YOLO has been successfully adapted for instance segmentation tasks, making it particularly useful for analyzing noise spectrograms in gravitational-wave data. In this context, YOLO can precisely segment and identify transient noise artifacts and potential astrophysical signals, providing detailed pixel-level masks that indicate the exact time-frequency coordinates of these events. This capability enhances the robustness and clarity of gravitational-wave analyses, enabling more effective noise mitigation and improved detection sensitivity.


The YOLO architecture is composed of three main components: the backbone, the neck, and the head. The backbone is responsible for extracting rich visual features from the input image. In YOLOv11~\citep{ultralytics_yolo}, this is typically based on a modified CSPDarknet or similar convolutional structure optimized for efficiency and depth \citep{wang2019cspnetnewbackboneenhance}. The neck, often implemented using PANet (Path Aggregation Network), further refines these features by fusing information across multiple scales, enhancing the model’s ability to detect objects of varying sizes \citep{liu2018pathaggregationnetworkinstance}. Finally, the head performs the core tasks of object detection and segmentation by predicting bounding boxes, class probabilities, and, in the case of YOLOv11 segmentation, pixel-level masks. This unified pipeline enables YOLO to perform real-time, high-accuracy segmentation.

\begin{figure*}[t]
  \centering

 {%
    \includegraphics[width=0.45\linewidth]{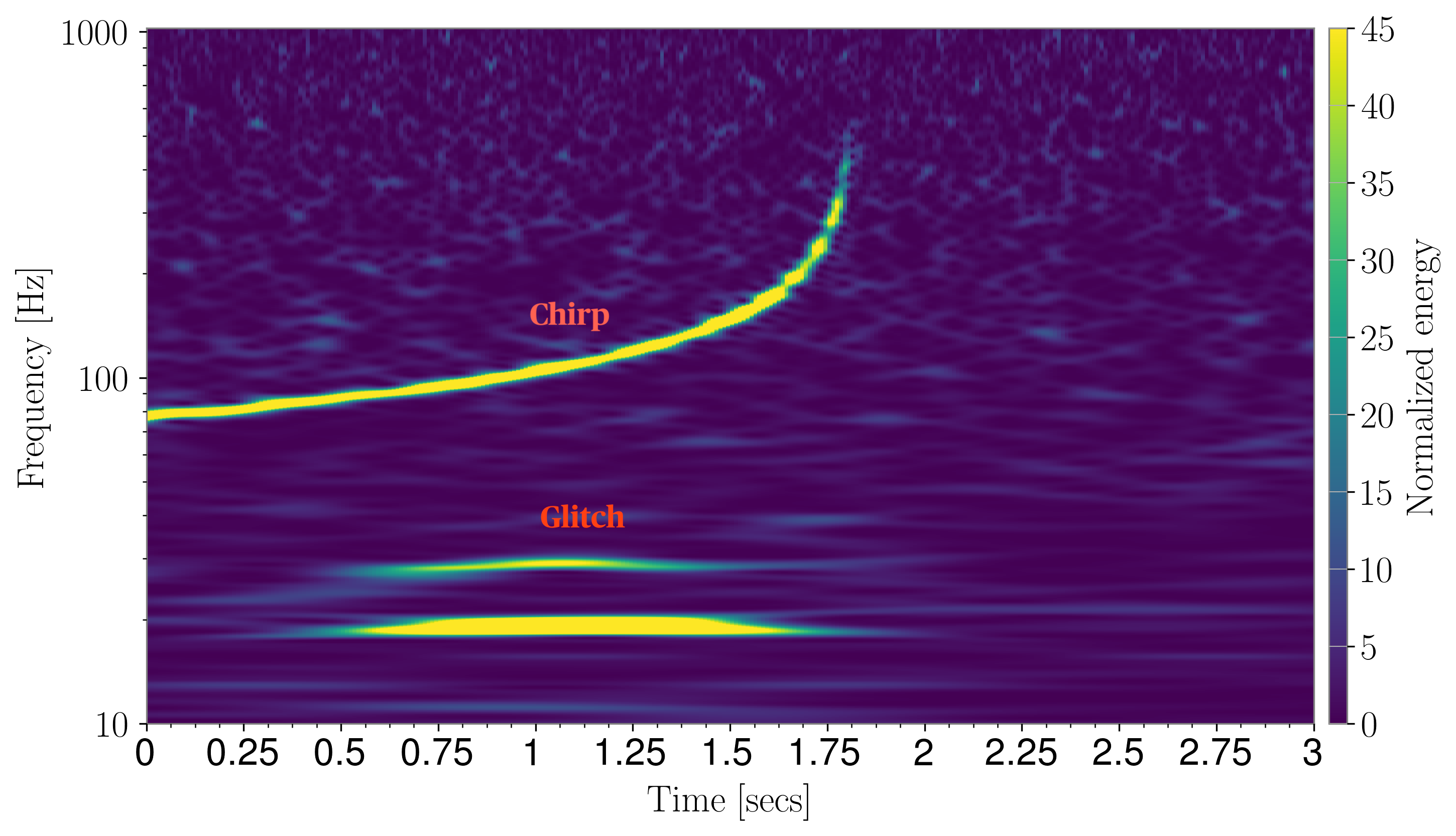}
    \label{fig:a}
  }
  {%
    \includegraphics[width=0.45\linewidth]{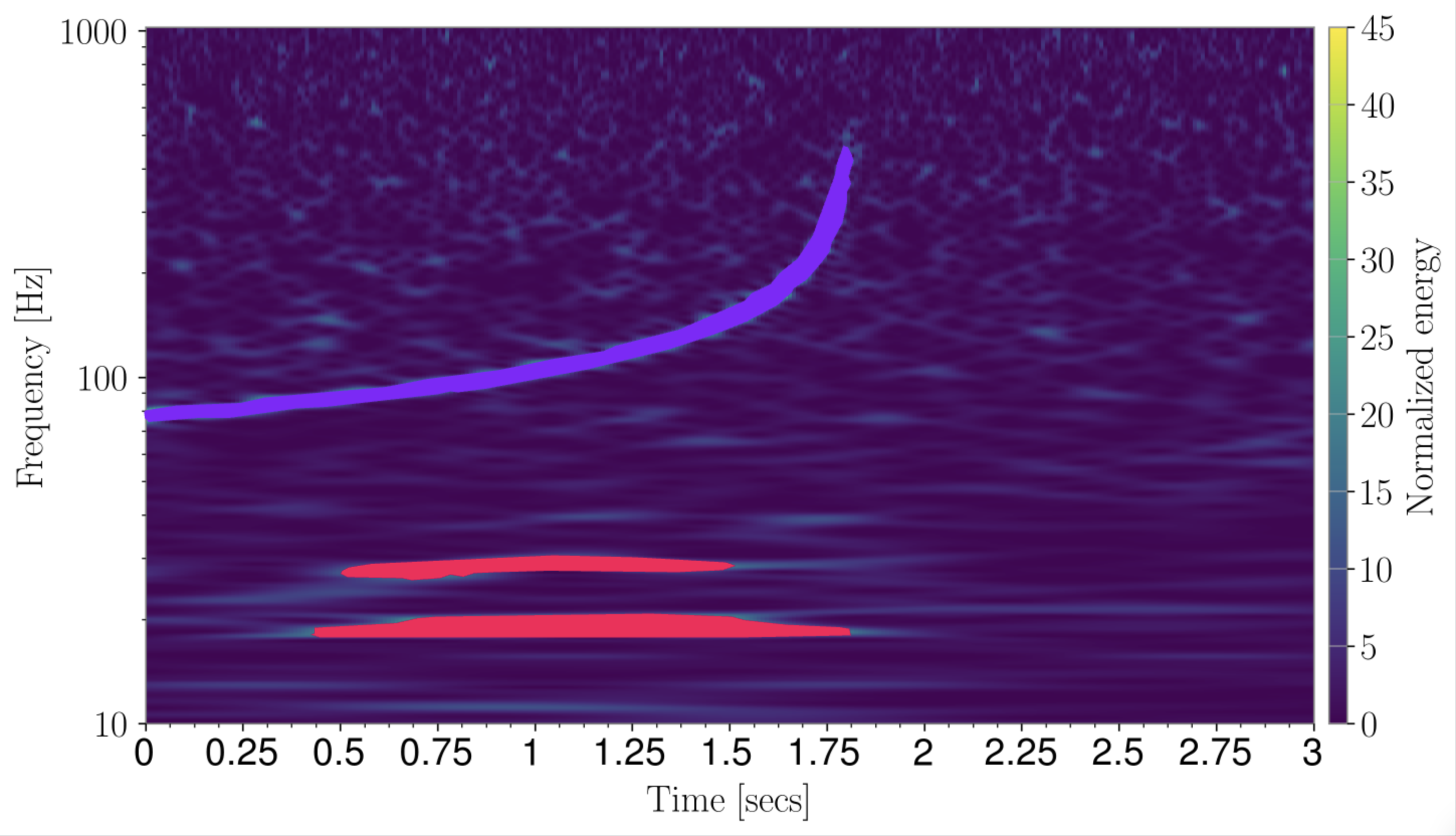}
    \label{fig:b}
  }

  \vspace{0.2cm}  

  {%
    \includegraphics[width=0.45\linewidth]{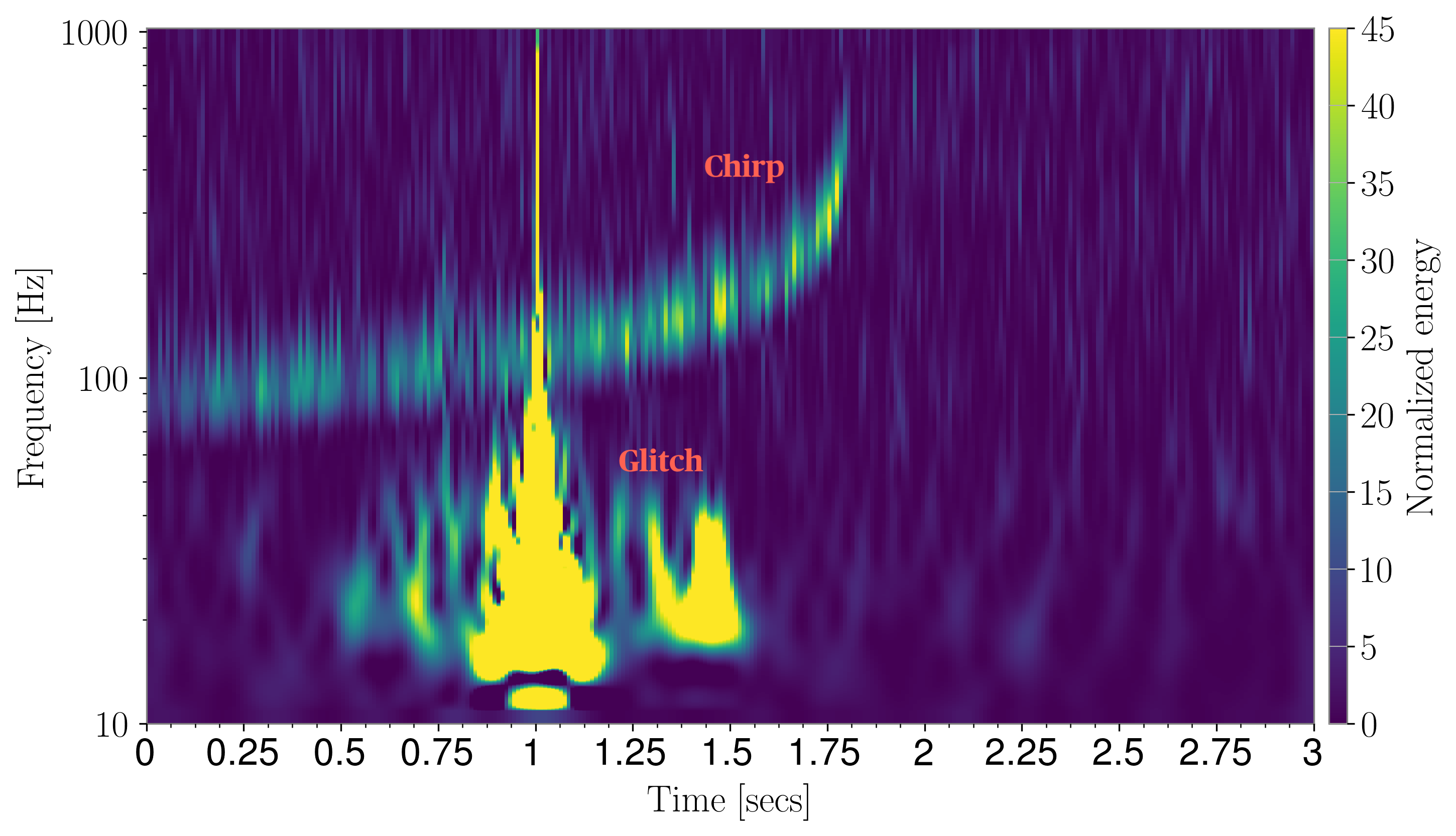}
    \label{fig:c}
  }
  {%
    \includegraphics[width=0.45\linewidth]{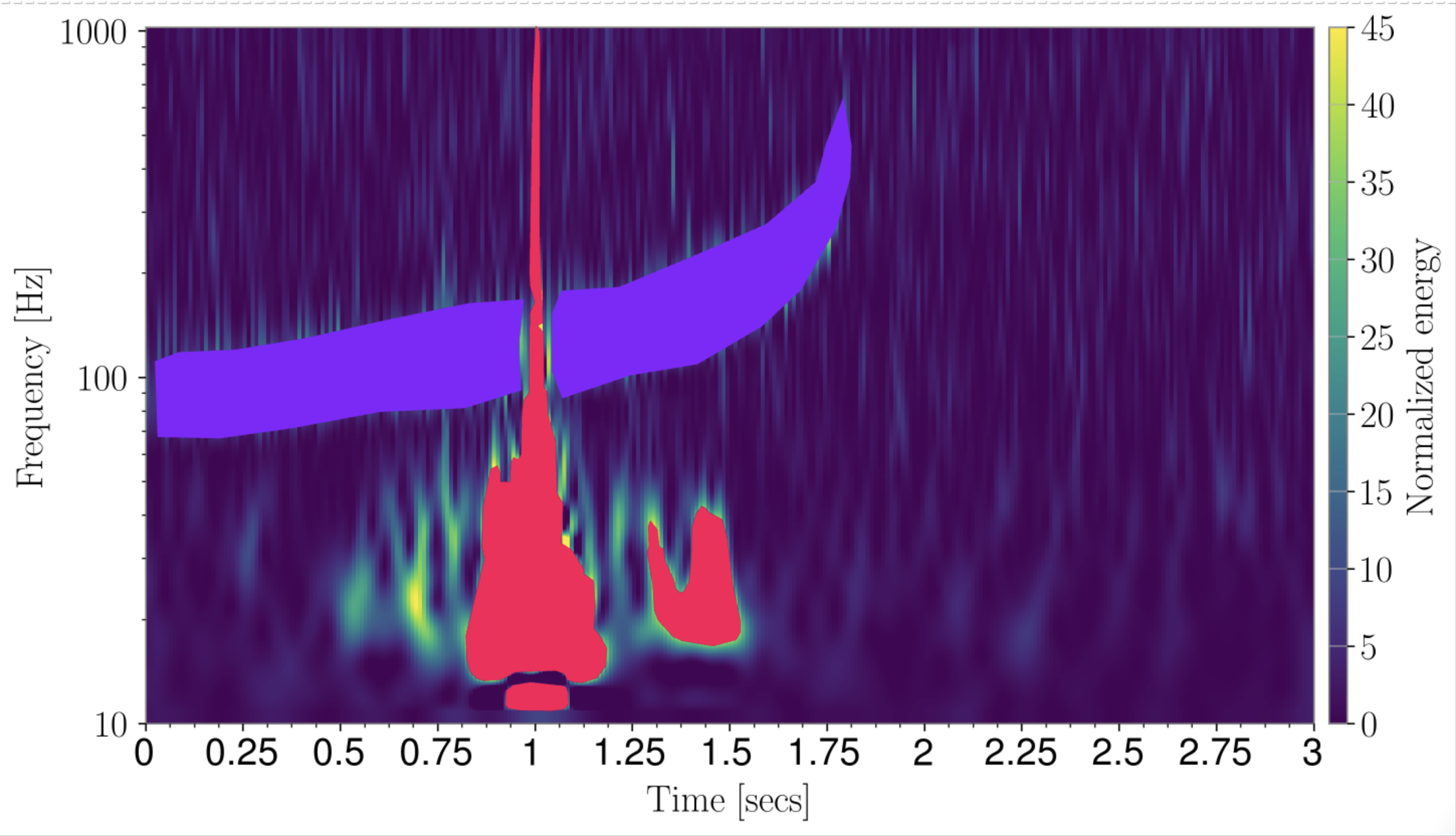}
    \label{fig:d}
  }

  \caption{\emph{Left}: Spectrograms (in the \emph{left} column panels) depicting both ``chirp" and ``glitch" signals present in the data. These are generated from the combined timeseries of chirps and glitches. These spectrograms are then annotated as shown in \emph{right} column panels with the labels corresponding to ``chirp" and ``noise". These annotated images form the dataset on which the segmentation model is then trained. }
    \label{fig:training_spectrograms}
\end{figure*}

\section{Training}

To train our YOLO-based segmentation model, we created a comprehensive
dataset by sampling transient noise (glitches) from LIGO's third
observing run (O3) GravitySpy dataset~\citep{glanzer2021gravity} and
generating binary black hole (BBH) and binary neutron star (BNS) chirp
signals using the PyCBC toolkit~\citep{usman2016_pycbc}. The GravitySpy
glitches were randomly sampled with SNR greater
than 7.5. The parameters for these synthetic chirp signals, including primary mass, secondary mass, and distance were sampled based on actual gravitational-wave detections from LIGO's first three observing runs (O1, O2, and O3) \citep{lvk_gwtc3_o3_sensitivity_2023}. 
For BNS we used $m_{p} < 2.5, m_s < 2.5$ and for BBH we have $m_{p} > 5, m_{s} > 5$, where $m_{p}$ and $m_{s}$ are the primary and secondary mass components respectively.
We combined glitch timeseries with chirp timeseries to form realistic datasets, from which spectrograms containing both noise and signals were produced \citep{gwpy_3_1_0}. Before initiating the training, we annotated the locations of glitches and chirp signals within these spectrograms and labeled them accordingly using the Roboflow platform, ensuring precise supervision for our segmentation model \citep{roboflow_software_2025}. Additionally, we sampled BNS signals from the O3 injection dataset~\citep{LIGO2021GWTC3} and followed the same procedure: combining glitch and chirp timeseries, creating spectrograms, and annotating the locations of glitches and signals before training. The annotations labels were ``chirp" and ``noise". 

To create the spectrograms, the strain data, originally sampled at 16,384~Hz, were resampled to
4096~Hz prior to computing the Q-transform. Q-transform spectrograms
were generated using a 3~s time window, a frequency range of
$10$--$1024~\mathrm{Hz}$, and a $Q$ range of $4$--$64$. The time and
frequency resolutions were set to $0.01~\mathrm{s}$ and
$0.05~\mathrm{Hz}$, respectively, and whitening was applied during
the Q-transform. The resulting spectrograms were rendered as images
and resized to $640\times640$ pixels for input to the YOLOv11 Nano
instance-segmentation model.
These spectrograms served as the training input for our model, enabling it to robustly distinguish astrophysical chirps from transient noise. Our final training dataset thus contains annotated spectrograms with simultaneous representations of glitches as well as both BBH and BNS signals. Fig. \ref{fig:training_spectrograms} shows the original and annotated spectrograms that were used for training. In these 3-second spectrograms, we varied the temporal separation between glitches and signals to enhance the model's training robustness.  

The training was carried out using a YOLOv11 Nano instance-segmentation model on the 
Roboflow hosted training framework \citep{roboflow_software_2025}. The final dataset consisted of 
722 annotated Q-transform spectrograms, divided into 578 training, 
72 validation, and 72 test images, corresponding to an 80/10/10 
split. This training dataset used in this work is publicly available on Zenodo
\citep{GW_dataset}. All input spectrograms were resized to $640\times640$ pixels 
using stretching, and no data augmentation was applied.
Training was
initialized from a previously trained model checkpoint
(\texttt{noise\_classification/17}) and continued for 100 epochs.
Training was 
configured for a maximum of 300 epochs and terminated after 100 epochs 
through the framework's early-stopping procedure. 
Performance was evaluated using standard metrics for segmentation models, including mean average precision at a 50{\%} intersection-over-union threshold (mAP@50), precision, and recall. 
Optimization 
hyperparameters, including the optimizer, batch size, and learning-rate 
schedule, were managed internally by the Roboflow training framework.
The trained model achieved an mAP@50 of $85.8\%$ on the
validation set and $88.2\%$ on the held-out test set.  The confidence
threshold that maximized the F1 score was 0.40, at which the
model achieved a precision of $92.9\%$, recall of $86.6\%$,
and F1 score of $89.6\%$.



\begin{align}
  \text{Precision}
    &= \frac{\text{True Positive}}{\text{True Positive} + \text{False Positive}} \label{prec},\\
  \text{Recall}
    &= \frac{\text{True Positive}}{\text{True Positive} + \text{False Negative}} \label{rec}.
\end{align}

The mAP@50 measures the model's ability to accurately segment and classify signals and glitches. And shown in Eq. ~\eqref{prec} and ~\eqref{rec}, precision measures the proportion of correct detections among all the positive predictions, and recall measures of all the actual positives, the fraction that was classified as positive. 
Another key metric is Confidence, which is a measure of how certain the model is that it has detected something within the bounding box parameters that belongs to a certain class. Furthermore, the F1 score given by Eq.~\eqref{F1} is a harmonic mean of the Precisian and Recall provides a measure of trade-off between the two.
These metrics collectively ensure a comprehensive evaluation of the YOLO model's accuracy and reliability in distinguishing signals and noise in gravitational-wave data. Table \ref{table:training_metrics} provides these metrics for the validation and test set. 

\begin{align}
  F_{1}
    &= \frac{2\,*\mathrm{Precision}*\mathrm{Recall}}
           {\mathrm{Precision} + \mathrm{Recall}}, \label{F1}
\end{align}




\begin{table}[!t]
\caption{Performance metrics for the validation and test sets.
Precision, recall, and F1 score are evaluated at a confidence
threshold of 0.40.}
\label{table:training_metrics}
\centering
\begin{tabular}{|c|c|c|c|c|}
\hline
Data Set & mAP50 & Precision & Recall & F1 Score \\
\hline
Validation & 85.8\% & 90.1\% & 82.4\% & 86.0\% \\
\hline
Test & 88.2\% & 92.9\% & 86.6\% & 89.6\% \\
\hline
\end{tabular}
\end{table}

\section{Inference}

To robustly evaluate the trained model on unseen data, we constructed four distinct inference datasets:

\begin{enumerate}
\item BBH chirps
\item BNS chirps
\item BBH chirps $+$ glitch
\item BNS chirps $+$ glitch
\end{enumerate}

The LIGO instrument data used in this study are coming from the detector site in Livingston, Louisiana.
The goal is to understand the performance of our model on chirps corresponding to binary black hole and binary neutron star coalescences, with and without the presence of glitches.
We did not generate signals corresponding to compact binaries made up of a black hole and a neutron star as this class of systems was not significantly adding to the diversity of chirp waveforms.
For the BBH chirps dataset, we randomly sampled injections from the O3 injection dataset used by the LIGO-Virgo-KAGRA collaborations in different signal-to-noise ratio (SNR) bands, between SNR values of 6 and 51 \citep{lvk_gwtc3_o3_sensitivity_2023}. Approximately 100 injections were selected in most SNR bands to ensure
diversity and to test model performance across a range of signal strengths.
The number of available injections decreases in the highest-SNR bins. Each BBH injection was then combined with a randomly sampled glitch from the O3 GravitySpy dataset to construct the BBH chirps $+$ glitch dataset.
A similar procedure was followed for BNS signals, yielding the BNS chirps and BNS chirps $+$ glitch datasets. The SNR range for this dataset is from 12 to 51. In total, each of the BBH datasets contains about 1400 samples, and each of the BNS datasets contains about 1100 samples.

\begin{figure*}[!t]
  \centering

 {%
    \includegraphics[width=0.45\linewidth]{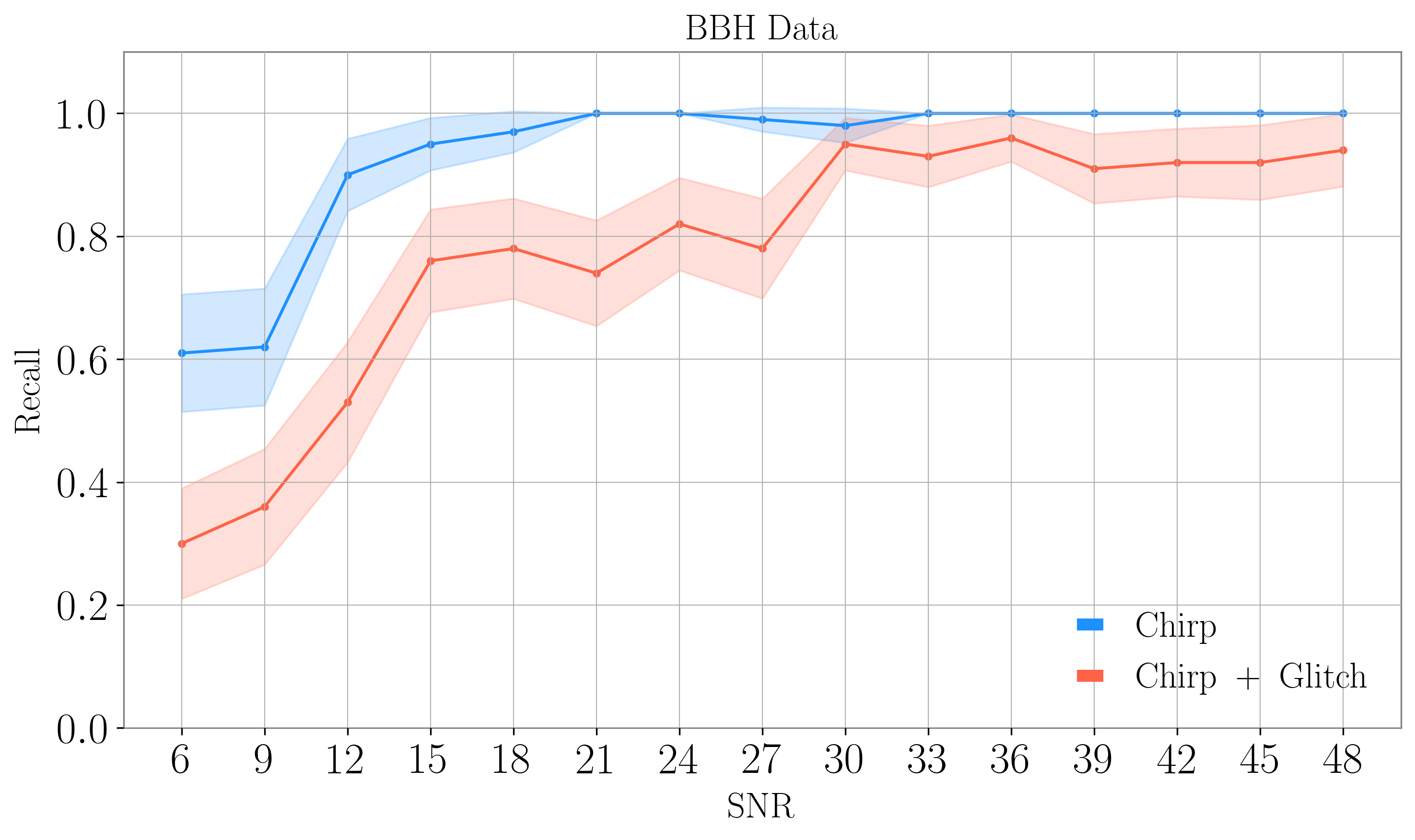}
    \label{fig:recall_bbh_chirps}
  }
  {%
    \includegraphics[width=0.45\linewidth]{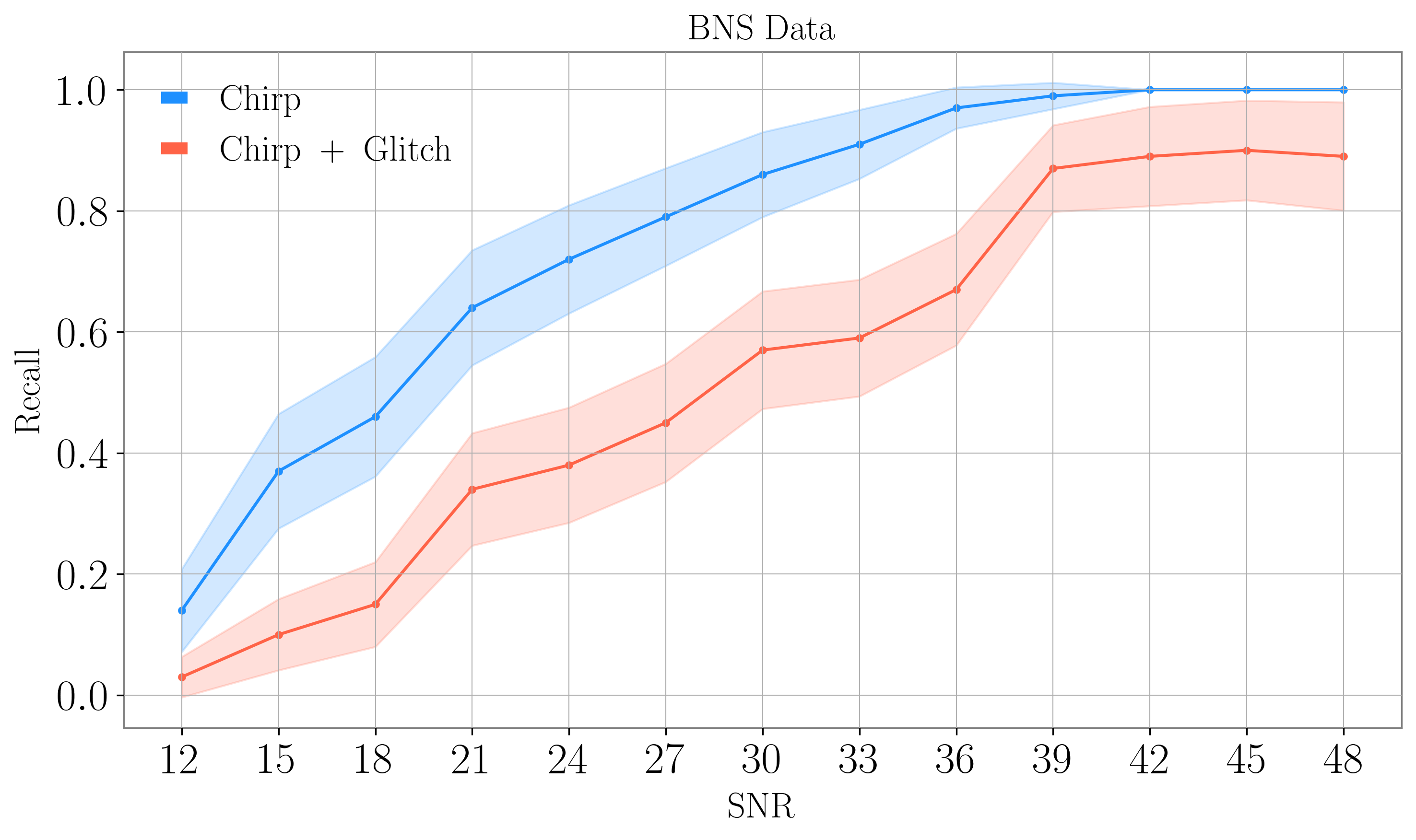}
    \label{fig:recall_bns_chirps}
  }
  \caption{Recall values for the detection of gravitational-wave chirp signals (BBH on \emph{left} plot and BNS on \emph{right}) with and without the addition of transient noise at model confidence value of 0.40. Each point represents a 3-wide SNR bin (e.g., 6--9, 9--12, 12--15). This is the value at which F1 score is maximized. The light shade band shows the $95\%$ confidence interval. As the SNR increases, the signals become more visible in the data, leading to higher recall values. Addition of transient noise adversely impacts the morphological appearance of the chirp signals in the data, thus causing a reduction in recall values. Fig. \ref{fig:bns_chirp_12_15}  and Fig. \ref{fig:chirp_scatter_noise} show and explain this concept in greater detail. Despite the addition of transient noise, the model is able to correctly identify a substantial fraction of chirps in the data in the mid to high SNR bands.}
    \label{fig:recall_vals}
\end{figure*}

Our goal is to evaluate how well the model performs across a wide range of SNR chirps, both in isolation and when accompanied by transient noise.
To simulate diverse interference scenarios, the temporal offset between the chirp event and the accompanying glitch was randomized, allowing the model to be tested under varying degrees of overlap and morphological disruption. The recall curves for all the four datasets are shown in Fig \ref{fig:recall_vals} at model's confidence value of 0.40. This is the confidence at which the F1 Score is maximized. More details on this are in the Appendix \ref{appendix_1}.

\section{Results}
\subsection{Chirps Dataset}\label{chirps_text}
\begin{figure*}[t]
  \centering

 {%
    \includegraphics[width=0.45\linewidth]{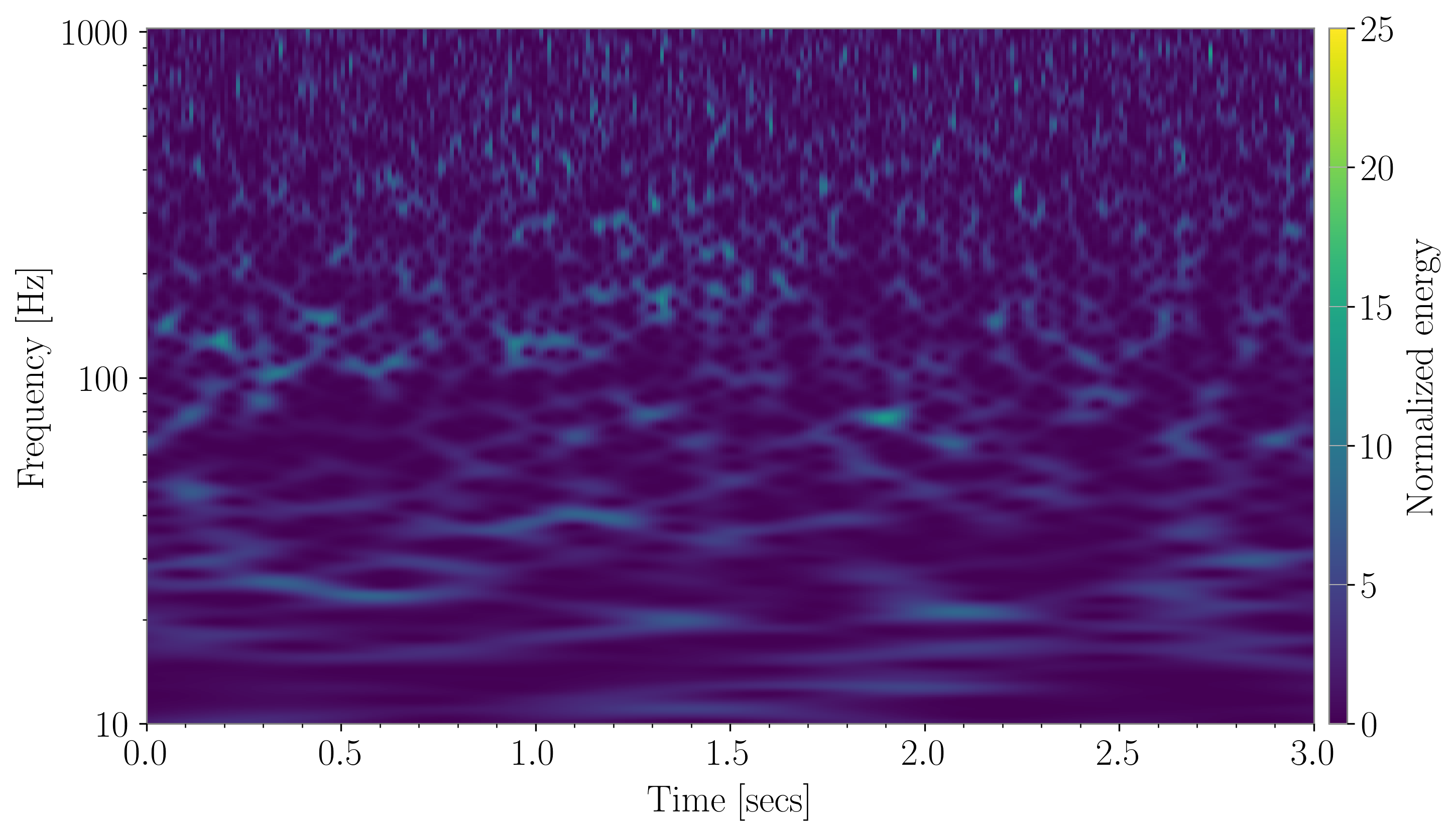}
    \label{fig:bns_chirp_snr12_15}
  }
  {%
    \includegraphics[width=0.45\linewidth]{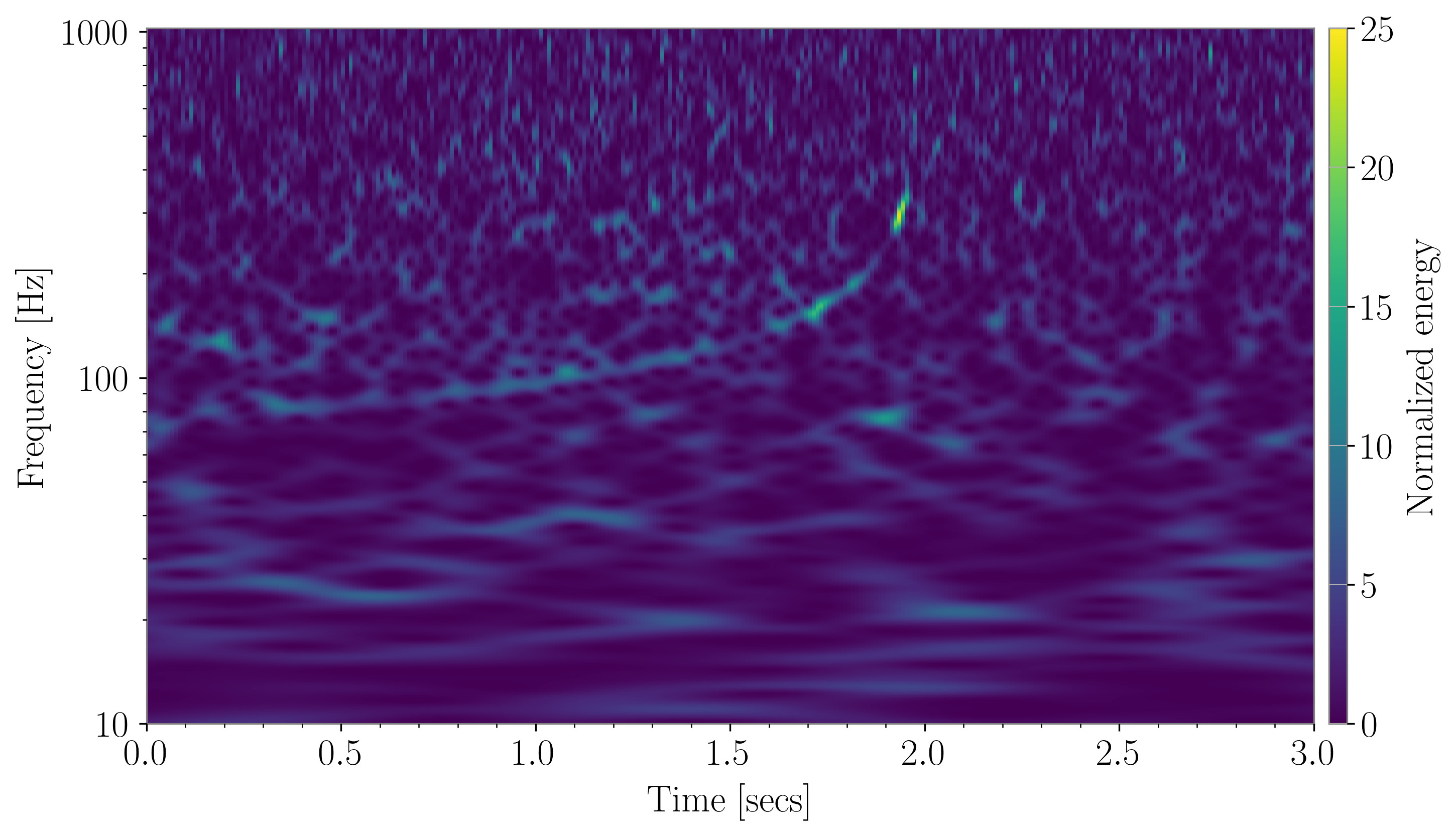}
    \label{fig:bns_chirp_snr12_15_iden}
  }
  \caption{Algorithmic performance on low SNR events. \emph{Left}: This spectrogram contains a BNS chirp in the SNR band $12 - 15$. Our segmentation model did not identify the presence of any ``chirp" in this image due to the absence of any morphological features resembling those of a chirp signal. \emph{Right}: This spectrogram is also based on a BNS chirp timeseries data in the same SNR band. Since a signal albeit a faint one showing a rise in frequency is visible here, the model was able to identify and localize the presence of ``chirp" in this data.  }
    \label{fig:bns_chirp_12_15}
\end{figure*}

Given the nature of gravitational-wave emission from compact binary systems, as the SNR of the chirp increases, the time-frequency features become increasingly pronounced in the Q-transforms. This leads to a larger proportion of chirps being identified by the model, as shown in Fig. \ref{fig:recall_vals}. At low SNRs, particularly for BNS chirps, the features are often too weak to appear distinctly in the Q-scans.
Fig \ref{fig:bns_chirp_12_15} shows two Q-scans of BNS chirp injections from the SNR band $12-15$. In the left image, no chirp morphology is visible, while in the right one a faint chirp can be seen. Consequently, the model successfully identified the chirp only in the right image. This limited visibility of chirp morphology in Q-scans at lower SNRs contributes to reduced recall performance in those regimes. As SNR increases, the chirp becomes more and more visible, leading to higher rate of identification. 

For BNS chirps in the right plot in Fig. \ref{fig:recall_vals}, the recall values show a consistent improvement with increasing SNR band. Starting from a recall of 0.14 in the 12--15 band, the performance gradually increases through the mid-SNR ranges and reaches 0.72 in the 24–27 band and then continues to grow. This trend reflects the model's increasing ability to correctly identify BNS chirps as the signal-to-noise ratio increases, which is expected given the enhanced visibility of chirp features at higher SNRs. Similarly, for BBH chirps in the left plot in Fig \ref{fig:recall_vals}, the model identifies more than 50\% of events even in the lowest SNR band, and achieves 90\% and up identification for signals above an SNR of 12.

\begin{figure*}[t]
  \centering

 {%
    \includegraphics[width=0.45\linewidth]{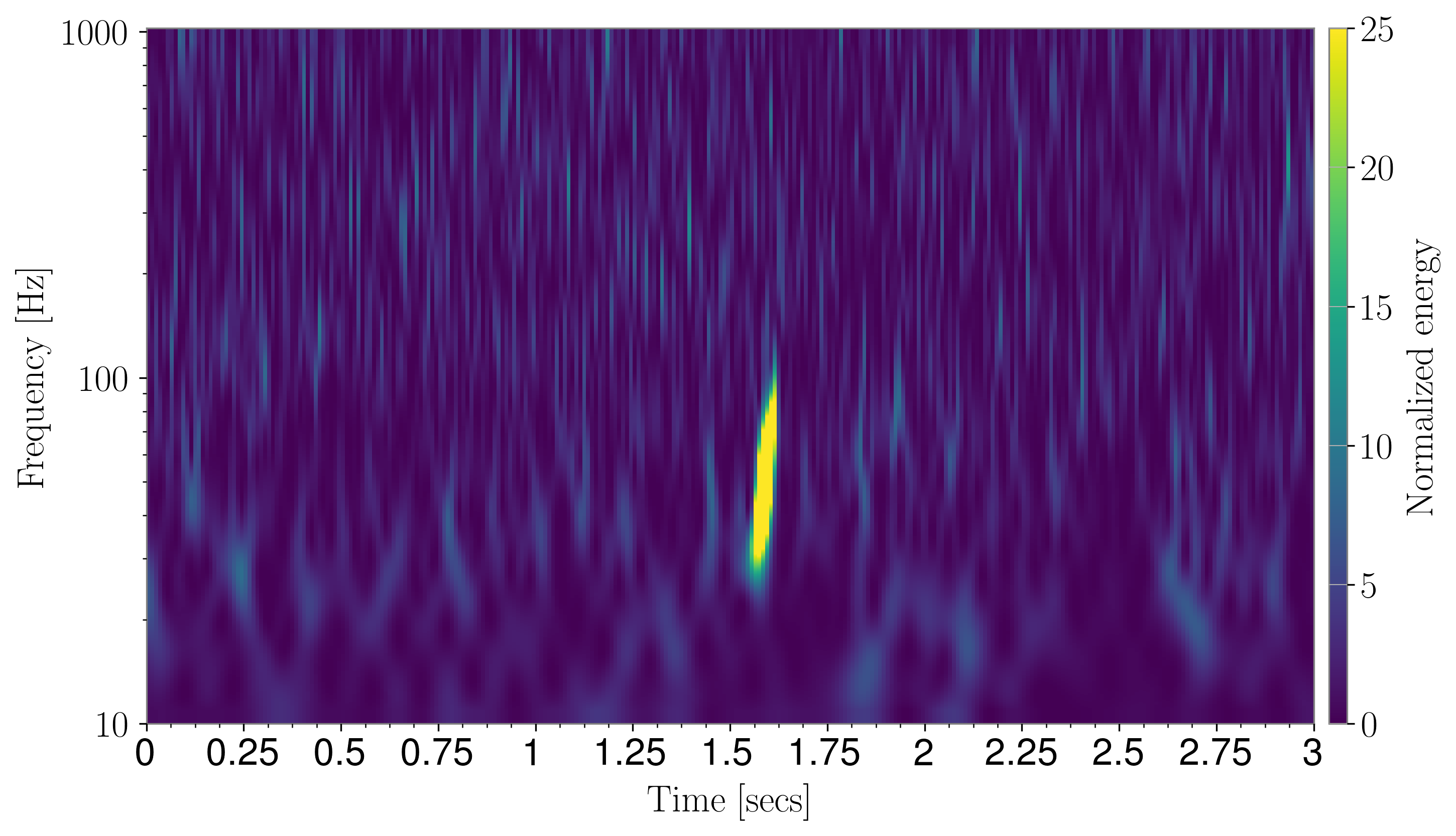}
    \label{fig:chirp_snr15}
  }
  {%
    \includegraphics[width=0.45\linewidth]{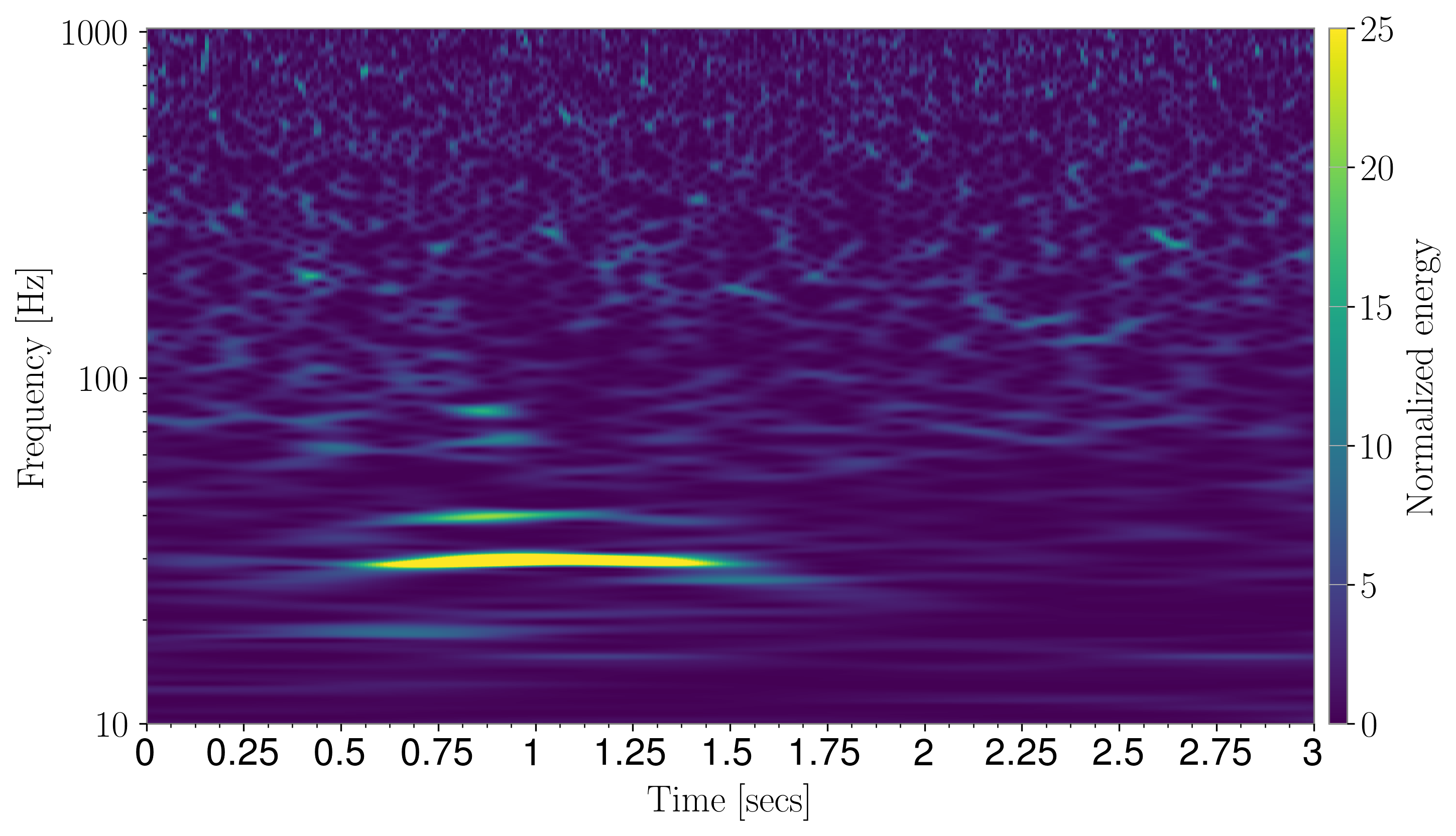}
    \label{fig:scatter_noise}
  }

  \vspace{0.2cm}  

  {%
    \includegraphics[width=0.45\linewidth]{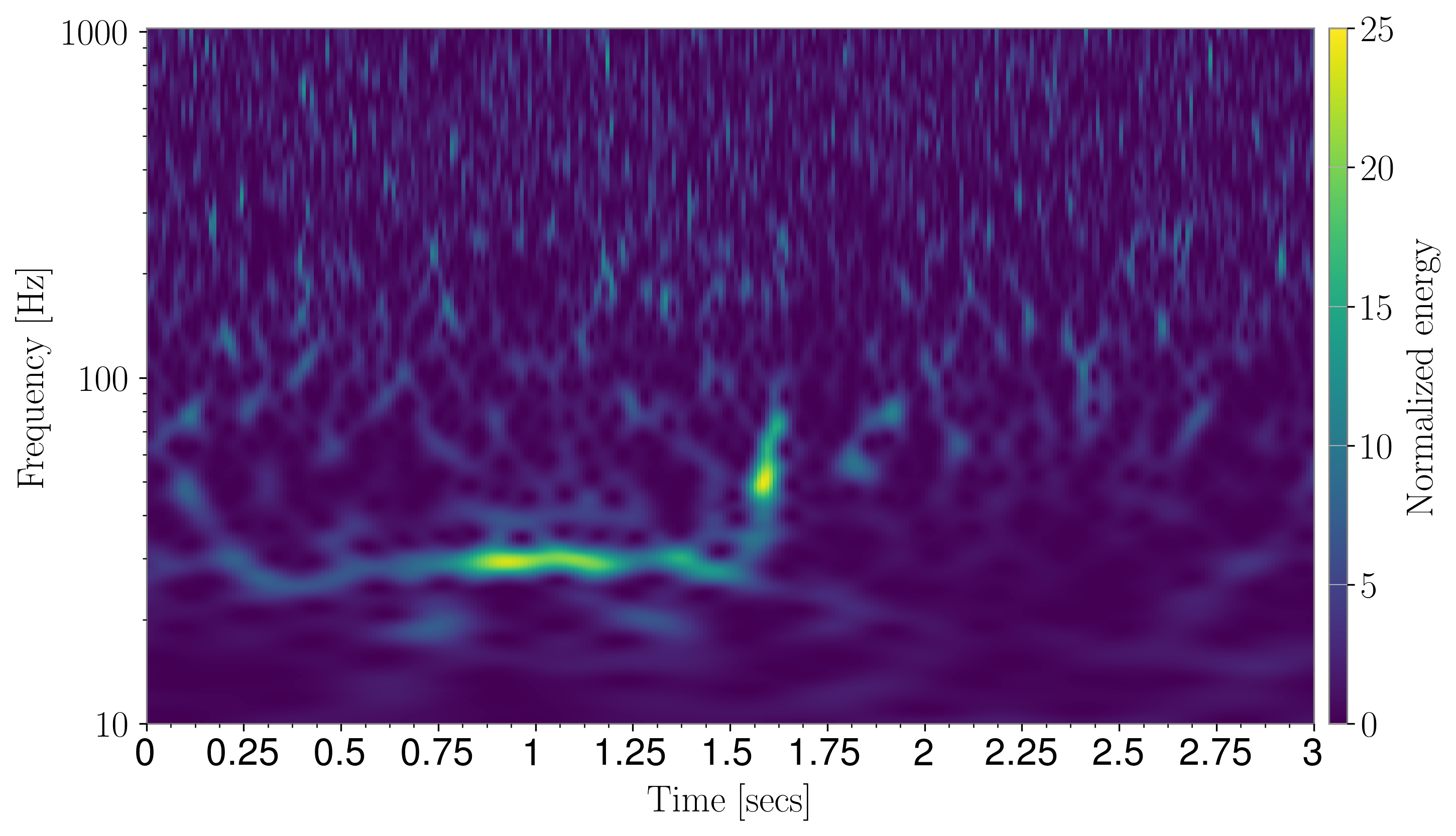}
    \label{fig:chirp_and_scatter}
  }
  \caption{This figure shows the morphological distortion that may happen when we combine two signals with different morphologies. The plot on the \emph{top left} shows a ``chirp" (BBH) signal while the plot on the \emph{top right} shows an instance of ``scattered light" glitch. When we combine these two signals, we obtain the plot on \emph{bottom} where neither features are easily visible. This can lead to misclassification by the segmentation model.}
    \label{fig:chirp_scatter_noise}
\end{figure*}
\subsection{Chirps + Glitch Dataset}\label{chirps_glitch_text}
As expected, the addition of glitches tends to reduce the model's recall rate. This reduction arises from the fact that glitches occurring in close temporal proximity to chirp signals can distort the underlying time-frequency morphology of the chirps in the spectrogram. Such distortions make it more difficult for the model to accurately recognize and segment the true astrophysical signal, thereby lowering its sensitivity. For instance, a BBH chirp that was initially recognized by the model may go undetected when overlaid with a transient glitch occurring in close temporal proximity. This is illustrated in fig \ref{fig:chirp_scatter_noise}. The top-left image shows an isolated BBH chirp, while the top-right one displays an isolated scattered light glitch. Both morphologies are clearly visible on their own. However, in the bottom image where the two are combined, the morphology of both the glitch and the BBH chirp is severely obscured, making it difficult for the model to identify either feature accurately.
This explains the reduction in the recall of chirps once the data is contaminated with transient noise as shown in Fig. \ref{fig:recall_vals}. 

Despite the anticipated drop in recall due to the addition of glitches, the model maintains a remarkably high level of performance. As shown in fig. \ref{fig:recall_vals}, the model continues to correctly identify a substantial fraction of chirps, even in the presence of transient noise. This robustness is particularly notable at higher SNR values, where the recall remains consistently high. The ability to accurately localize and segment chirps despite the morphological distortions introduced by glitches highlights the effectiveness and resilience of the proposed YOLO-based approach.

\begin{figure*}[t]
  \centering

 {%
    \includegraphics[width=0.45\linewidth]{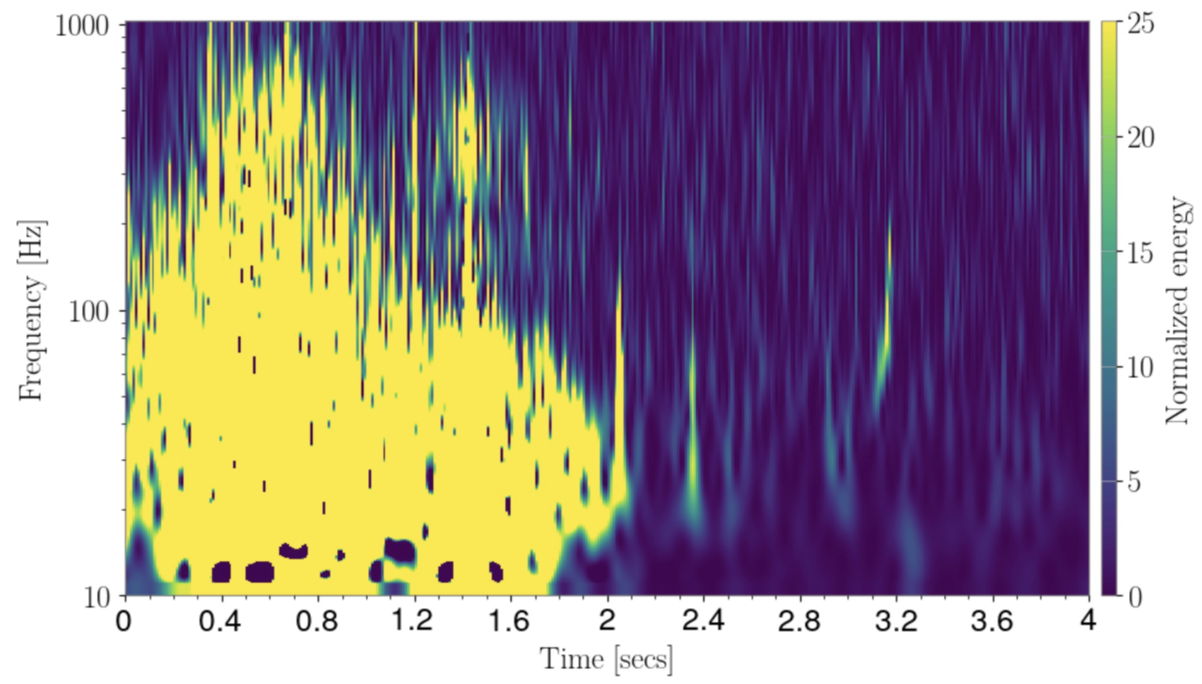}
    \label{fig:bbh_loud_}
  }
  {%
    \includegraphics[width=0.45\linewidth]{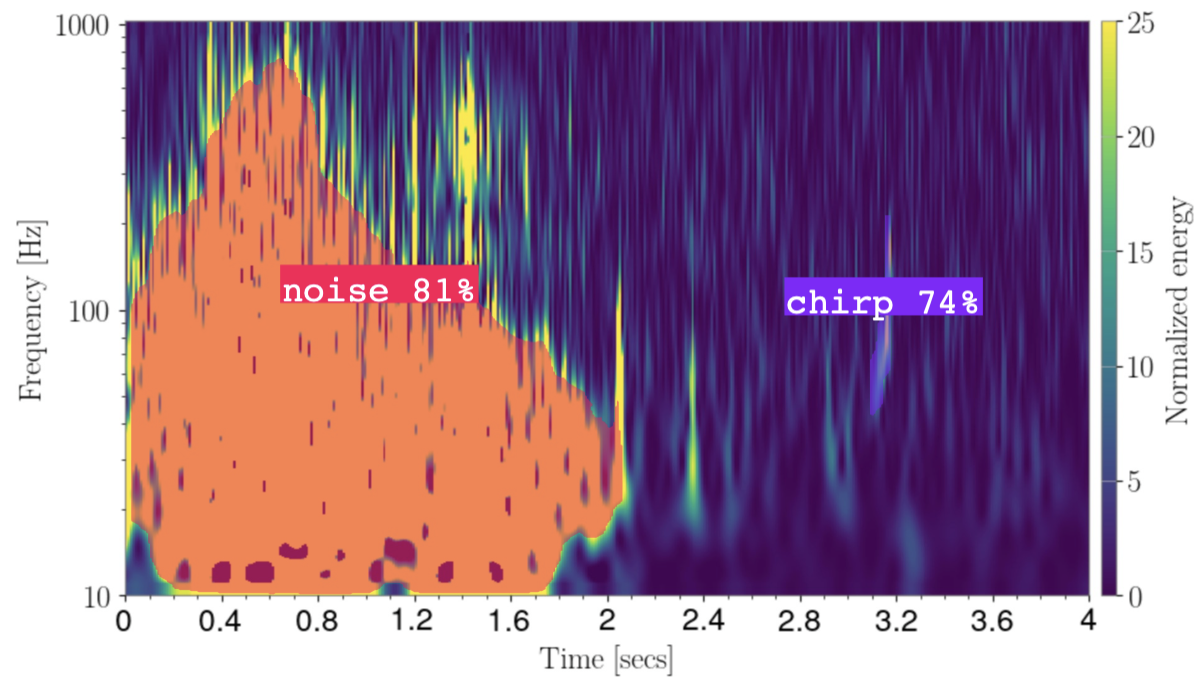}
    \label{fig:bbh_loud_m}
  }

  \vspace{0.2cm}  

  {%
    \includegraphics[width=0.45\linewidth]{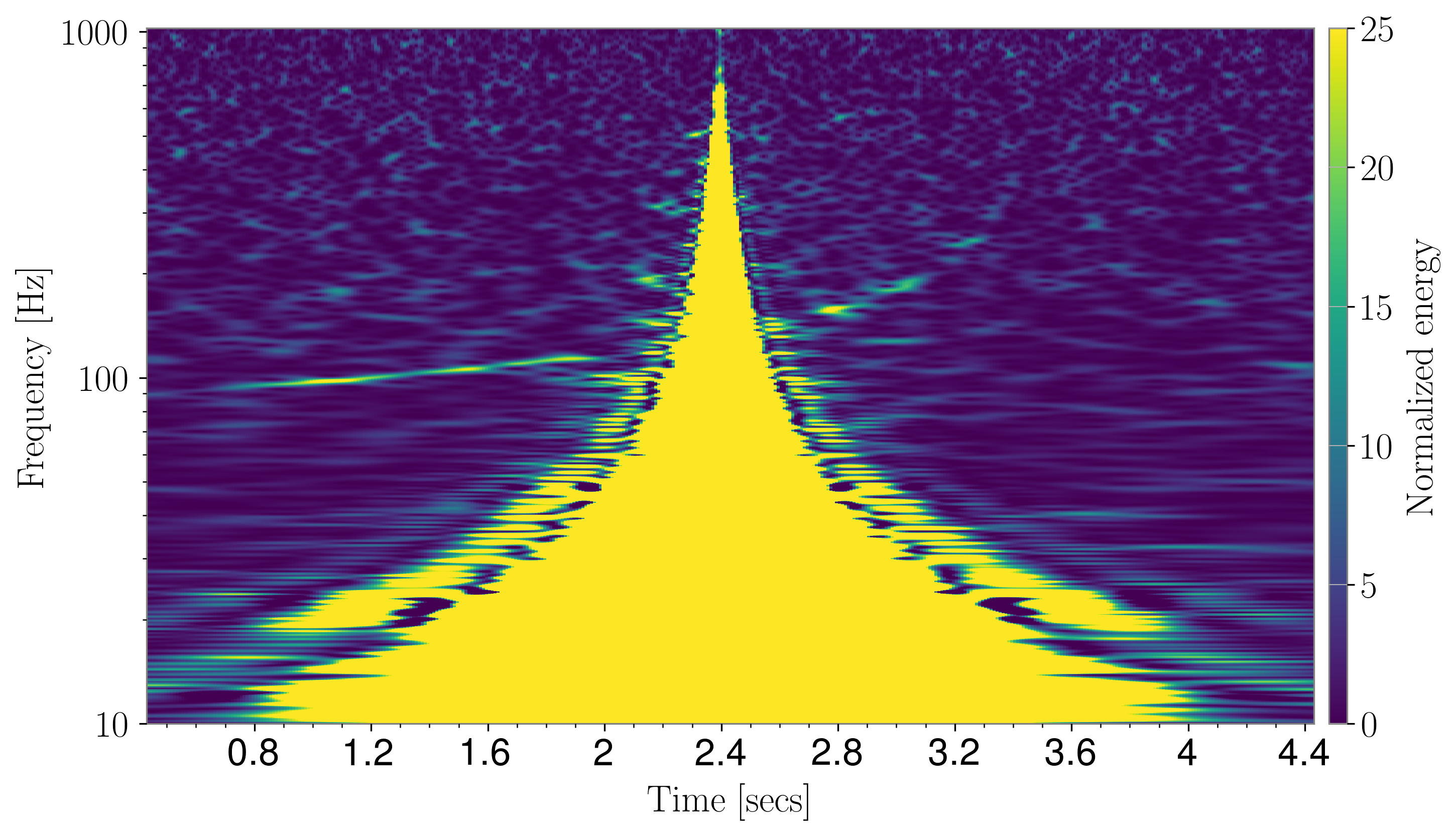}
    \label{fig:bns_loud_}
  }
  {%
    \includegraphics[width=0.45\linewidth]{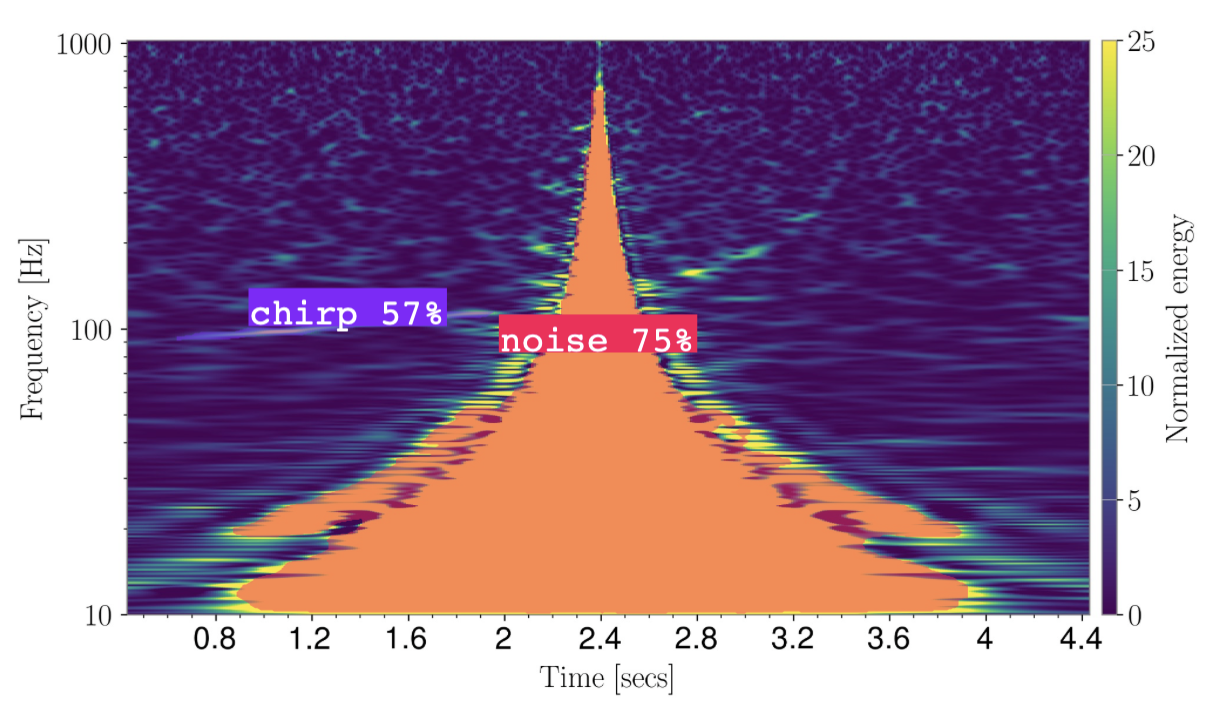}
    \label{fig:bns_loud_m}
  }

  \caption{Plots showing the inference on data containing both chirp signals and transient noise. \emph{Top left}: The Q-transform shows a very high SNR glitch in the first half of the duration window and a chirp signal following close to 3.2 sec mark. \emph{Top right}: The segmentation model identifies and localizes the presence of noise and signal with a confidence of $81$\% and $74$\% respectively. \emph{Bottom left}: This image shows the BNS event GW170817 at LIGO Livingston where a loud glitch overlaps with the astrophysical chirp signal. \emph{Bottom right}: The model correctly detects the presence of both the noise and chirp signal in the data with a confidence of $75 \%$ and $57 \%$ respectively. These examples demonstrate the model's ability to identify the presence of chirp signals even when very loud transient events are present in the vicinity.}
    \label{fig:chirps_loud_}
\end{figure*}

This is shown in Fig \ref{fig:chirps_loud_}, the top and botton left plots show a very high SNR glitch in close proximity to BBH and BNS chirp respectively. The plots on the right shows that the model correctly identified the chirps with high enough confidence in these Q-scans. 

Importantly, this capability extends current gravitational-wave data-analysis approaches by enabling the identification of chirps in the immediate vicinity of general transient noise. To our knowledge, this scenario has not previously been systematically investigated using an image-segmentation framework.
This simultaneous detection of astrophysical signals amidst real, unmodeled glitches is a novel contribution, and highlights the unique strength of our YOLO-based segmentation framework in handling complicated time-frequency interference scenarios.

\begin{figure}[tb]
  \centering
  \includegraphics[width=0.5\textwidth, height=0.23\textwidth]{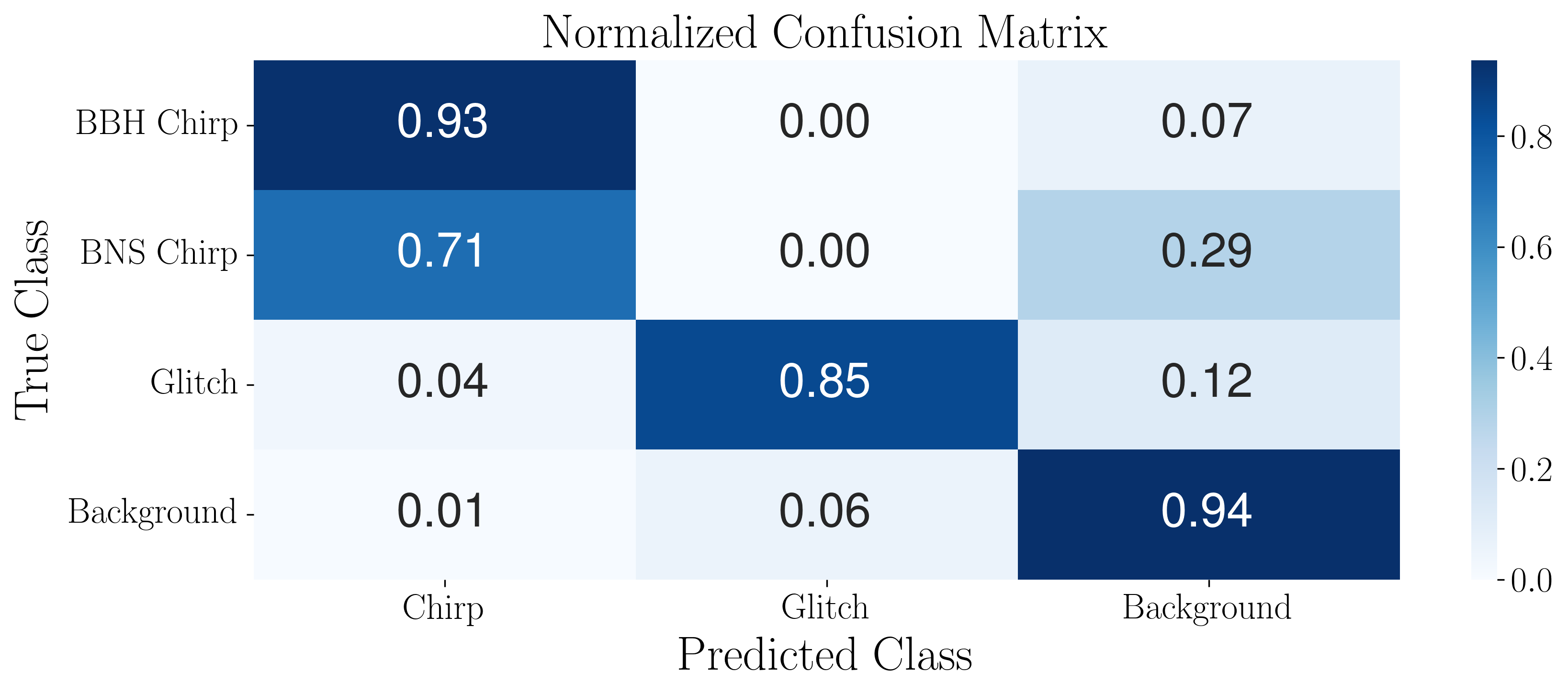}
  \caption{Normalized confusion matrix showing model performance across BBH chirp, BNS chirp, glitch, and background classes. BBH chirps and glitches are classified with high accuracy, while BNS chirps show some confusion with background.}
  \label{fig:conf_matrix}
\end{figure}

\subsection{Confusion Matrix}
The analysis in 
\ref{chirps_text} and 
\ref{chirps_glitch_text} explores the model’s ability to identify chirps both in the absence and presence of glitches as the signal-to-noise ratio varies. Beyond detection performance, we also aim to understand the general classification behavior of the model when presented with chirps, glitches and background. To this end, we performed inference on four datasets: BBH chirps, BNS chirps, a glitch dataset and a quiet background dataset. The glitch dataset consists of approximately 1400 randomly sampled glitches from the O3 Livingston data, each with an SNR greater than 7.5 and the quite background dataset consists of about 400 instances where we do not expect any transient events. The results of this classification evaluation are summarized in the confusion matrix shown in Fig.~ \ref{fig:conf_matrix}.

We have discussed the segmentation analysis on BBH chirps and BNS chirps in the section \ref{chirps_text} where the chirps were divided into different SNR bands. The overall true positive rate of BBH chirps and BNS chirps as shown in the Fig. \ref{fig:conf_matrix} tracks with the results shown in Fig. \ref{fig:recall_vals}. Additionally, this confusion matrix shows that our segmentation model does a
good job at identifying just transient noise and just background as well. This is important because sometimes search pipelines can generate alerts on transient noise; what the confusion matrix shows is that our segmentation model can quickly diagnose such an event as ``glitch" with high probability \citep{KAGRA:2021vkt}. 

\section{Event Validation with GW-YOLO}\label{ev_yolo}
Following the identification of a gravitational-wave candidate, the LIGO-Virgo-KAGRA collaborations undertake several data quality checks are performed on the event. This procedure is referred to as ``event validation'' \citep{LIGO:2024kkz}. The event validation report contains data quality details around the time of an event, including the (manual) identification of the presence of any transient noise near the event time. The latter is typically quantified in the form of noise box coordinates. 
These coordinates identify the region in the spectrogram where the transient noise occurs. These time-frequency noise box values are then used for noise mitigation purposes, including guiding algorithms for effective noise subtraction \citep{Cornish:2014kda}. Currently, this estimation of time-frequency noise box parameters is done manually by the event validation volunteers through visual inspection. With the segmentation tool we are presenting here, this process can not only be automated but also made more precise since the model provides pixel-level masks outlining the presence of transient noise. With the ever increasing rate of detections and decreasing person power for the manual data quality checks, this automation is becoming essential. 

\begin{figure}
    \centering
    \includegraphics[width=\linewidth]{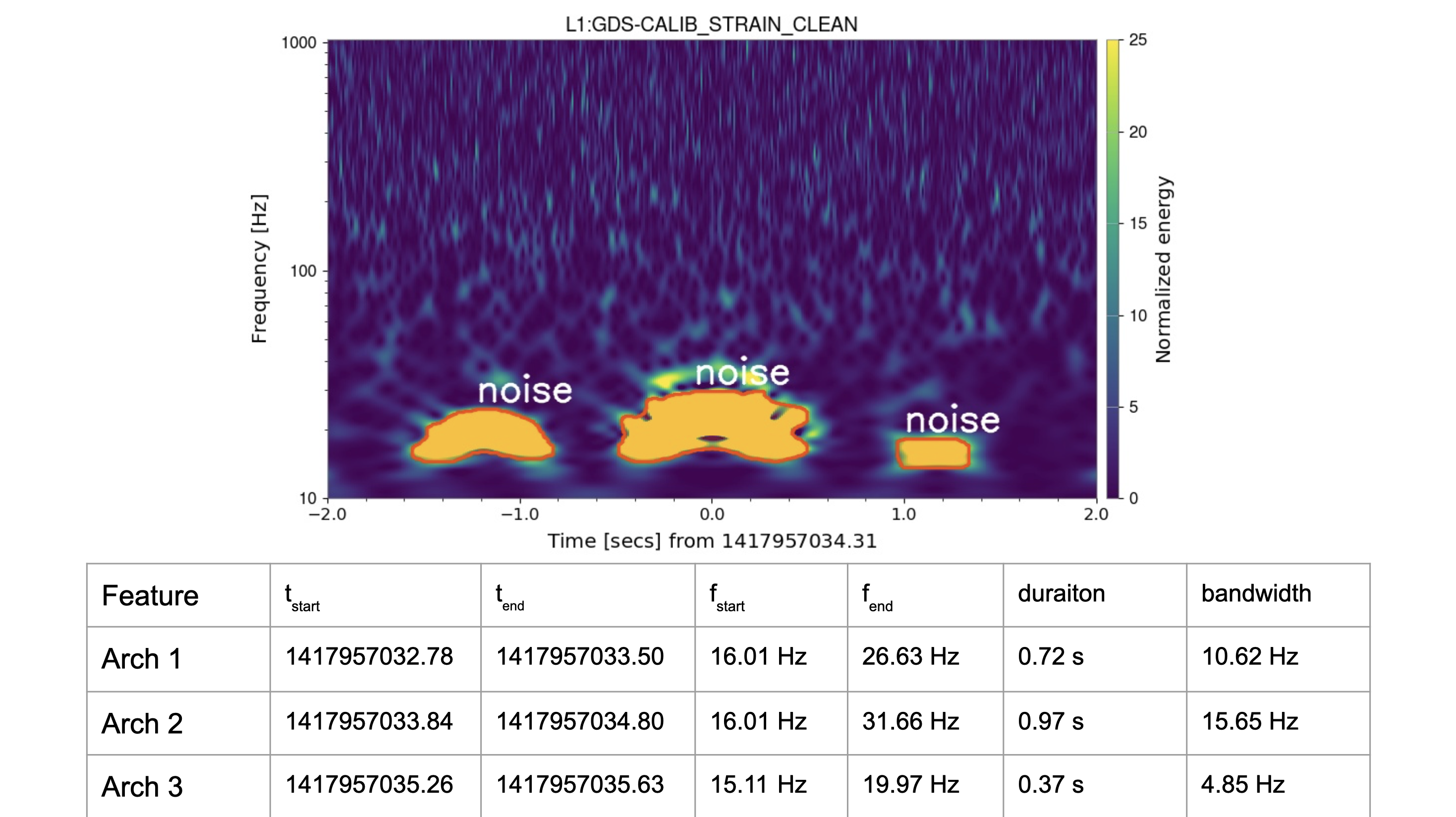}
    \caption{GW-YOLO segmentation and corresponding time–frequency characterization of transient noise. Top: Q-transform of LIGO strain data showing three transient-noise arches identified and segmented by GW-YOLO. Bottom: Time–frequency properties of the detected arches derived from the GW-YOLO JavaScript Object Notation (JSON) output using the accompanying analysis utilities. For each segmented feature, the start and end GPS times, lower and upper frequency bounds, duration, and bandwidth are reported. The arches are ordered by their start time.}
    \label{fig:yolo_noise_class}
\end{figure}

In addition to the visual segmentation output, GW-YOLO produces a machine-readable JSON data product for each analyzed event. The JSON file records the event metadata and the properties of each detected instance, including its predicted class, confidence score, bounding box, and the pixel-level coordinates defining the segmentation mask. These masks provide a compact representation of the time-frequency morphology identified by GW-YOLO and can be processed directly without relying on the annotated Q-scan visualization. The analysis utilities provided with GW-YOLO map the pixel-space coordinates of each detected instance onto the physical coordinates of the Q-transform, providing the start and end GPS times and frequency bounds of each segmented feature. From these quantities, the duration and frequency bandwidth of individual GW signals or transient-noise features are also obtained. This is shown in Fig \ref{fig:yolo_noise_class}. The JSON output can therefore serve as a machine-readable data-quality product containing the information required for automated event validation and integration with downstream low-latency analysis pipelines.

\begin{table}[t]
\centering
\caption{Latency of the GW-YOLO event-validation pipeline, measured over 50 events on the LIGO Data Grid node. The total latency corresponds to the time from the start of processing to the availability of the segmentation output.}
\label{tab:gwyolo_latency}
\begin{tabular}{lcc}
\hline
\textbf{Pipeline stage} & \textbf{Mean (s)} & \textbf{Median (s)} \\
\hline
Data retrieval + Q-transform       & 1.511  & 1.476 \\
Q-scan image generation            & 7.916  & 7.882 \\
YOLO inference + postprocessing    & 1.178  & 1.157 \\
\hline
\textbf{Total latency} & \textbf{10.604} & \textbf{10.482} \\
\hline
\end{tabular}
\end{table}

\subsection{Latency results}

We additionally evaluated the end-to-end latency of the GW-YOLO inference pipeline to assess its suitability for low-latency event validation. The latency was measured over 50 independent events on an LIGO Data Grid computing node. Starting from a GPS event time, retrieving the strain data and computing its Q-transform required a mean latency of 1.51 s, while generating the Q-scan used as input to the segmentation model required 7.92 s. Model inference and subsequent processing required an additional 1.18 s, resulting in a mean latency of 10.60 s from the start of processing to the availability of the segmentation output (median 10.48 s). The dominant contribution to the overall latency arises from rendering and saving the Q-transform as an image rather than from the segmentation model itself. This indicates that substantial reductions in latency may be achievable through optimization of the spectrogram-generation stage without modification of the trained GW-YOLO model.

\section{Discussion}

Our findings highlight the potential of YOLO-based neural network architectures in gravitational-wave detection, particularly in managing the complexities introduced by transient noise. By reframing the problem within a computer vision context, as the one of scene identification, we have shown that this approach can address challenges that existing methods often struggle with. For instance, it can effectively handle situations where astrophysical signal detection based on matched filtering is misled by
glitches or where probabilistic classifiers have difficulty with overlapping events.
As gravitational-wave detector sensitivity continues to improve, we anticipate an increased rate of astrophysical sources being detected by the instruments, including longer-duration signals from binary neutron star mergers, and possibly signals influenced by gravitational lensing. 
Both the rate and the time-frequency characteristics of future gravitational-wave detections increase the likelihood of multiple, and often overlapping, transient events appearing within a single analysis window. Addressing these complexities will require further innovation, and our approach offers a promising step in that direction. 

Interpreting the predictions of neural networks is inherently challenging. However, YOLO's ability to provide pixel-level localization significantly improves interpretability, eliminating the need for additional techniques such as class activation mapping~\citep{zhou2016learning}. Moreover, the precise time-frequency localization obtained from our scheme can be leveraged in downstream tasks, such as identifying and cleaning glitch-contaminated segments.
Handling very faint Q-scan tracks from low-SNR events remains a challenge for our scheme. We plan to address this in future work, potentially by improving the preprocessing and data preparation stages—specifically through adaptive selection of Q-scan parameters optimized to pickup chirp-like patterns in the time-frequency maps.

\begin{figure}[!t]
  \centering
  \includegraphics[width=0.95\linewidth]{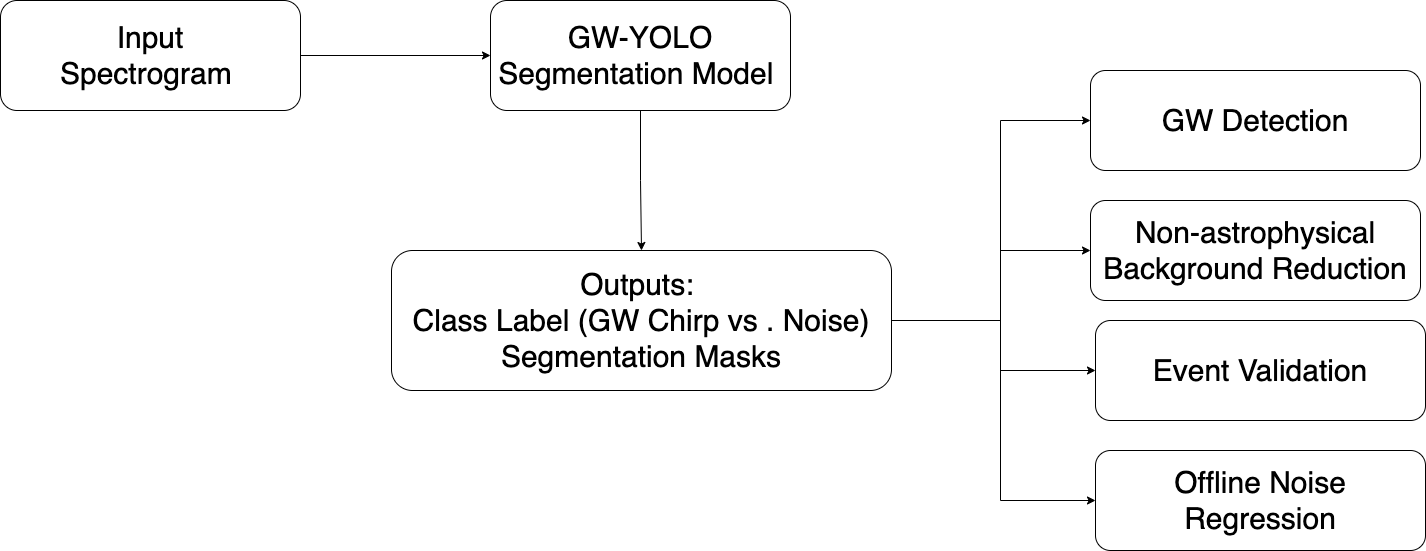}
  \caption{Overview of the proposed GW–YOLO workflow. A Q-transform spectrogram (Q-scan) is processed by a segmentation model that outputs a class label (GW chirp vs.\ noise) and time–frequency masks. This paper evaluated detection in the presence of overlapping events and glitches; the other prospective downstream applications include—non-astrophysical background reduction, low-latency event validation, and offline noise regression.}
  \label{fig:single}
\end{figure}

To contextualize these results and to outline near-term extensions, we summarize the intended GW–YOLO workflow in Fig.~\ref{fig:single}. We aim to deploy the trained model to continuously process Q-scans and return (i) a class label (GW chirp vs.\ noise) and (ii) time–frequency segmentation masks that localize the predicted instances. In this paper, we used these outputs solely for detection under realistic scenarios with overlapping events and non-astrophysical glitches, and quantified the performance via an injection study. The right-hand blocks in Fig.~\ref{fig:single} depict downstream applications enabled by the same outputs but left for future work: rapid GW detection; reduction of non-astrophysical background through mask-informed vetoes; event validation that can lessen human-in-the-loop latency; and offline noise regression/deglitching by flagging or excising contaminated time–frequency regions.

Starting in 2015, gravitational-wave detectors provide astronomical alerts in low-latency for the purpose of following up the astrophysical sources via electromagnetic wave and neutrino observations in real-time. This necessitates the real-time processing of strain data and corresponding noise identification and mitigation with signal detection algorithms running on dedicated computing hardware. YOLO has been widely adopted in domains like autonomous driving and as such it offers a natural fit for deployment in LIGO's real-time detection pipelines. 
Finally, current event validation efforts within LIGO still rely heavily on human-in-the-loop, which can introduce latency, subjectivity and will not scale well with the expected increase in detection rates. 
Our proposed approach provides a natural means of reducing reliance on manual event validation by offering a more objective, automated system that can scale efficiently, ultimately supporting a more robust and reproducible analysis framework.

\section{Data Release}
\label{sec:data_availability}

The training dataset used in this work is publicly available through
Zenodo~\citep{GW_dataset}. The trained models associated with this work
will be made publicly available upon acceptance of the manuscript.

\appendix
\section{Segmentation Model Metrics during Training}\label{appendix_1}
\begin{figure*}[t]
    \centering
    \includegraphics[width=0.8\linewidth]{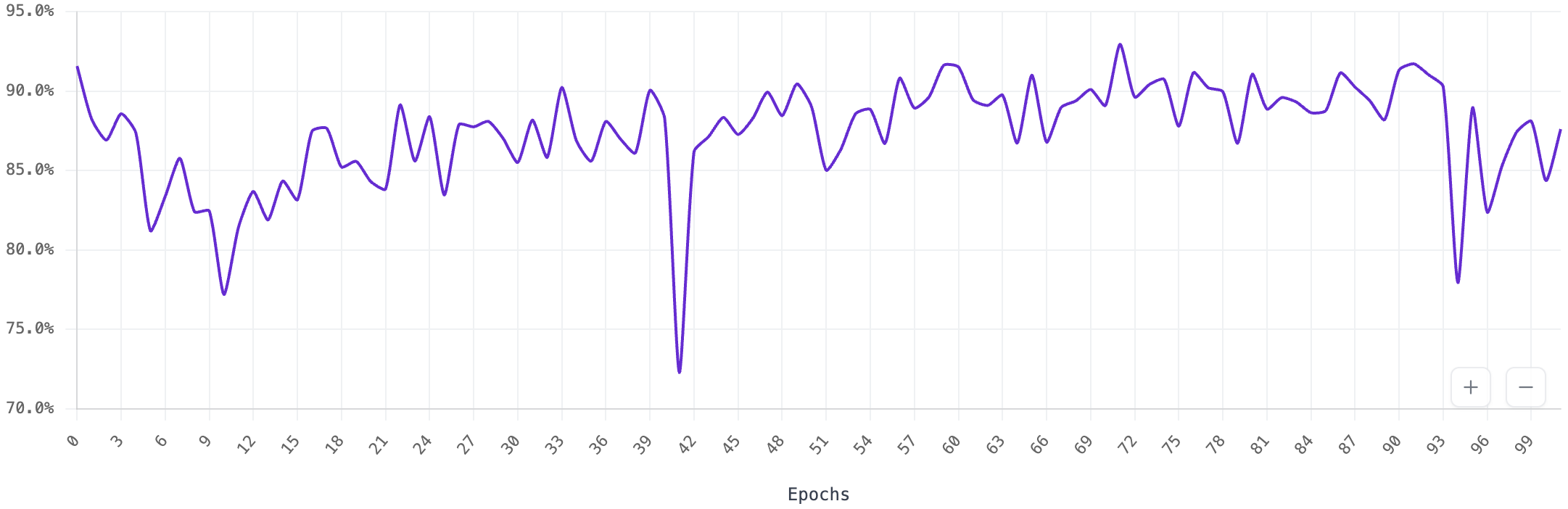}
    \caption{Mean average precision (mAP) as a function of training epoch for the YOLOv11 Nano instance-segmentation model \citep{roboflow_software_2025}. Training was initialized from the previously trained \texttt{noise\_classification/17} checkpoint, resulting in high mAP from the beginning of the training run. The model was subsequently fine-tuned for 100 epochs using the Roboflow hosted training framework. }
    \label{fig:training_map}
\end{figure*}

\begin{figure*}[t]
  \centering

 {%
    \includegraphics[width=0.8\linewidth]{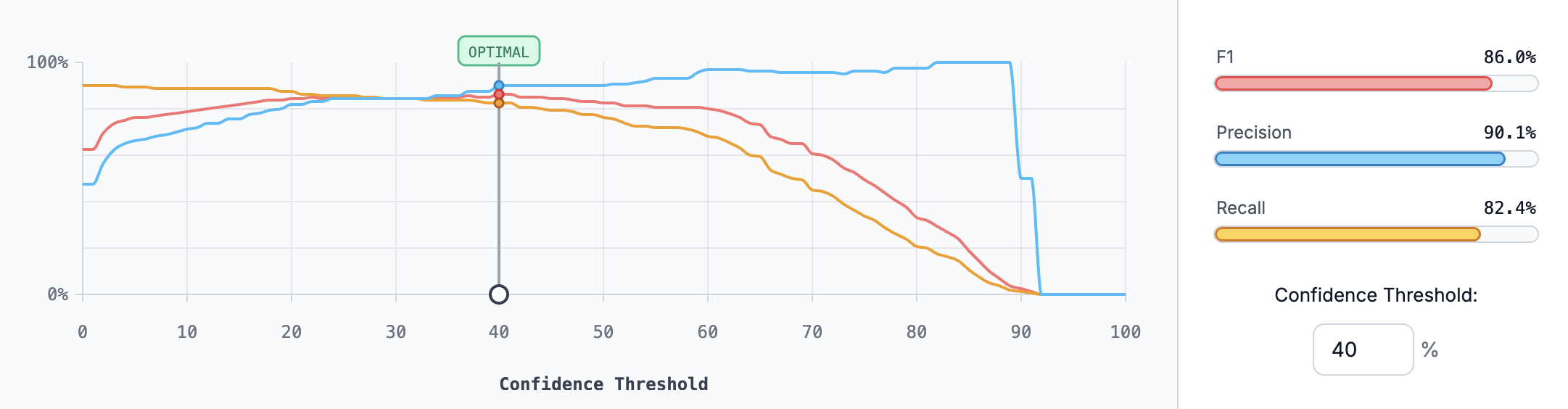}
    \label{fig:valid_prf1}
  }
  \vspace{0.2em}
  {%
    \includegraphics[width=0.8\linewidth]{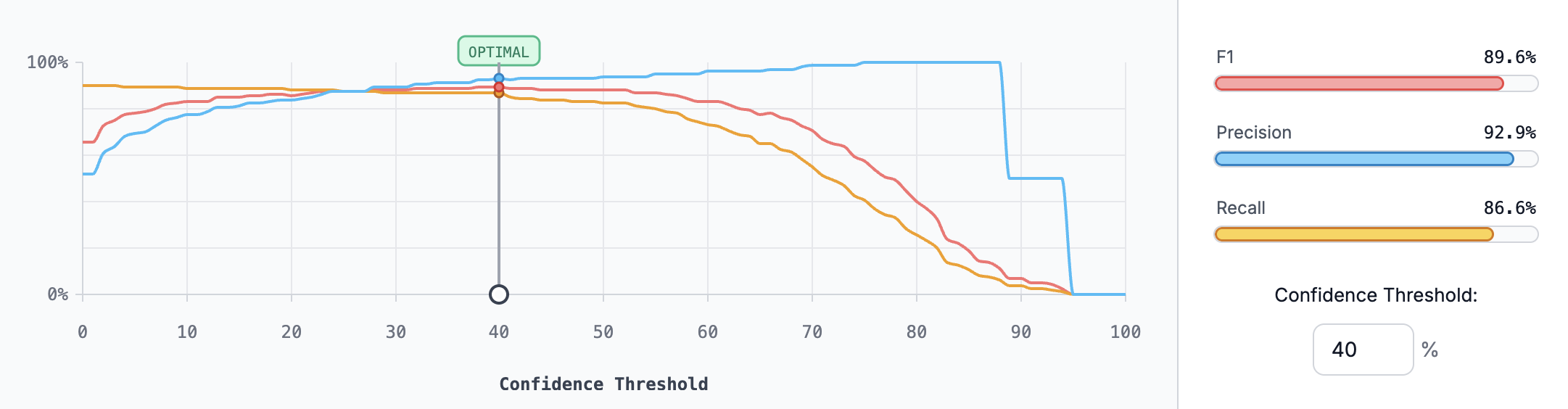}
    \label{fig:test_prf1}
  }
  \caption{Precision, recall, and F1 score as functions of the confidence
threshold for the validation (top) and test (bottom) sets, as evaluated
using the Roboflow model evaluation framework \citep{roboflow_software_2025}. The F1-optimal confidence
threshold is 0.40 for both datasets. At this threshold, the model achieves
precision, recall, and F1 scores of 90.1\%, 82.4\%, and 86.0\%,
respectively, on the validation set, and 92.9\%, 86.6\%, and 89.6\%,
respectively, on the test set.} 
    \label{fig:valid_test}
\end{figure*}

The final GW-YOLO model was obtained through iterative refinement of the training dataset. The final training run was initialized from the (\texttt{noise\_classification/17}) checkpoint and fine-tuned using the 722-image dataset described in Section 4. Figure \ref{fig:training_map} shows the mAP evolution during this final training run, while Figure \ref{fig:valid_test} shows the confidence-dependent precision, recall, and F1 scores on the validation and held-out test sets.

These metrics such as mean Average Precision (mAP) over epochs, precision–recall curves, and F1 score curves, collectively illustrate the model's performance progression and class-wise detection quality. These graphs help assess the trade-offs between precision and recall, convergence behavior, and the effectiveness of incremental dataset refinement through iterative annotation and retraining. 




\newpage
\begin{acknowledgments}
SS, NM and EK acknowledge support from the National Science Foundation (NSF) under award PHY-1764464 and PHY-2309200 to the LIGO Laboratory. NM acknowledges support from the National Science Foundation under Cooperative Agreement PHY-2019786 (The NSF AI Institute for Artificial Intelligence and Fundamental Interactions, http://iaifi.org/) and from MathWorks, Inc. Also, EK acknowledges support from the National Science Foundation under PHY-2117997 (The NSF Institute on Accelerated AI Algorithms for Data Driven Discovery--A3D3, http://a3d3.ai/). We also thank Tom Dooney for the LIGO-Virgo-KAGRA internal pre-publication review of the paper. This research was undertaken with the support of the LIGO computational clusters.
This material is based upon work supported by NSF's LIGO Laboratory which is a major facility fully funded by the National Science Foundation.

\end{acknowledgments}


\bibliography{references}

@article{PhysRevD.91.082001,
  title = {Gravitational wave detector with cosmological reach},
  author = {Dwyer, Sheila and Sigg, Daniel and Ballmer, Stefan W. and Barsotti, Lisa and Mavalvala, Nergis and Evans, Matthew},
  journal = {Phys. Rev. D},
  volume = {91},
  issue = {8},
  pages = {082001},
  numpages = {6},
  year = {2015},
  month = {Apr},
  publisher = {American Physical Society},
  doi = {10.1103/PhysRevD.91.082001},
  url = {https://link.aps.org/doi/10.1103/PhysRevD.91.082001}
}

@article{evans2021horizon,
  title={A horizon study for cosmic explorer: science, observatories, and community},
  author={Evans, Matthew and Adhikari, Rana X and Afle, Chaitanya and Ballmer, Stefan W and Biscoveanu, Sylvia and Borhanian, Ssohrab and Brown, Duncan A and Chen, Yanbei and Eisenstein, Robert and Gruson, Alexandra and others},
  journal={arXiv preprint arXiv:2109.09882},
  year={2021}
}

@article{punturo2010einstein,
  title={The Einstein Telescope: A third-generation gravitational wave observatory},
  author={Punturo, M and Abernathy, Matt and Acernese, Fausto and Allen, Bruce and Andersson, Nils and Arun, K and Barone, Fabrizio and Barr, Bryan and Barsuglia, Matteo and Beker, Mark and others},
  journal={Classical and Quantum Gravity},
  volume={27},
  number={19},
  pages={194002},
  year={2010},
  publisher={IOP Publishing}
}

@article{abbott2017gw170817,
  author    = {B. P. Abbott and R. Abbott and T. D. Abbott and F. Acernese and K. Ackley and C. Adams and T. Adams and P. Addesso and R. X. Adhikari and V. B. Adya and others},
  title     = {GW170817: Observation of Gravitational Waves from a Binary Neutron Star Inspiral},
  journal   = {Physical Review Letters},
  volume    = {119},
  number    = {16},
  pages     = {161101},
  year      = {2017},
  doi       = {10.1103/PhysRevLett.119.161101}
}

@article{biswas2013application,
  author    = {Rahul Biswas and Lindy Blackburn and Junwei Cao and Reed Essick and Kari Alison Hodge and Erotokritos Katsavounidis and Kyungmin Kim and Young-Min Kim and Eric-Olivier Le Bigot and Chang-Hwan Lee and John J. Oh and Sang Hoon Oh and Edwin J. Son and Ye Tao and Ruslan Vaulin and Xiaoge Wang},
  title     = {Application of Machine Learning Algorithms to the Study of Noise Artifacts in Gravitational-Wave Data},
  journal   = {Physical Review D},
  volume    = {88},
  number    = {6},
  pages     = {062003},
  year      = {2013},
  publisher = {American Physical Society},
  doi       = {10.1103/PhysRevD.88.062003},
  url       = {https://link.aps.org/doi/10.1103/PhysRevD.88.062003}
}

@article{powell2017classification,
  author    = {Jade Powell and Alejandro Torres-Forné and Ryan Lynch and Daniele Trifirò and Elena Cuoco and Marco Cavaglià and Ik Siong Heng and José A. Font},
  title     = {Classification methods for noise transients in advanced gravitational-wave detectors II: performance tests on Advanced LIGO data},
  journal   = {Classical and Quantum Gravity},
  volume    = {34},
  number    = {3},
  pages     = {034002},
  year      = {2017},
  publisher = {IOP Publishing},
  doi       = {10.1088/1361-6382/34/3/034002},
  url       = {https://dx.doi.org/10.1088/1361-6382/34/3/034002}
}

@article{George2018DeepFiltering,
  author  = {Daniel George and E. A. Huerta},
  title   = {Deep Learning for Real-Time Gravitational Wave Detection and Parameter Estimation: Results with Advanced LIGO Data},
  journal = {Physics Letters B},
  volume  = {778},
  pages   = {64--70},
  year    = {2018},
  doi     = {10.1016/j.physletb.2017.12.053}
}

@article{Gabbard2018Matching,
  author  = {Hunter Gabbard and Michael Williams and Fergus Hayes and Chris Messenger},
  title   = {Matching Matched Filtering with Deep Networks for Gravitational-Wave Astronomy},
  journal = {Physical Review Letters},
  volume  = {120},
  number  = {14},
  pages   = {141103},
  year    = {2018},
  doi     = {10.1103/PhysRevLett.120.141103}
}

@article{Mukund2017Transient,
  author  = {N. Mukund and S. Abraham and S. Kandhasamy and S. Mitra and N. S. Philip},
  title   = {Transient Classification in LIGO Data Using Difference Boosting Neural Network},
  journal = {Physical Review D},
  volume  = {95},
  number  = {10},
  pages   = {104059},
  year    = {2017},
  doi     = {10.1103/PhysRevD.95.104059}
}

@article{Zevin2017GravitySpy,
  author  = {Michael Zevin and others},
  title   = {Gravity Spy: Integrating Advanced LIGO Detector Characterization, Machine Learning, and Citizen Science},
  journal = {Classical and Quantum Gravity},
  volume  = {34},
  number  = {6},
  pages   = {064003},
  year    = {2017},
  doi     = {10.1088/1361-6382/aa5cea}
}

@article{Razzano2018ImageBased,
  author  = {Massimiliano Razzano and Elena Cuoco},
  title   = {Image-Based Deep Learning for Classification of Noise Transients in Gravitational Wave Detectors},
  journal = {Classical and Quantum Gravity},
  volume  = {35},
  number  = {9},
  pages   = {095016},
  year    = {2018},
  doi     = {10.1088/1361-6382/aab793}
}

@article{Coughlin2019Discovering,
  author  = {Scott B. Coughlin and others},
  title   = {Classifying the Unknown: Discovering Novel Gravitational-Wave Detector Glitches Using Similarity Learning},
  journal = {Physical Review D},
  volume  = {99},
  number  = {8},
  pages   = {082002},
  year    = {2019},
  doi     = {10.1103/PhysRevD.99.082002}
}

@techreport{LIGO:P2500369,
  author       = {Man Leong Chan and others},
  title        = {GSpyNetTreeS : a machine learning solution for glitch localization in time and frequency},
  number       = {LIGO-P123456},
  year         = {2025},
  url          = {https://dcc.ligo.org/LIGO-P2500369},
}

@inproceedings{Redmon2016YOLO,
  author    = {Joseph Redmon and Santosh Divvala and Ross Girshick and Ali Farhadi},
  title     = {You Only Look Once: Unified, Real-Time Object Detection},
  booktitle = {Proc.\ IEEE Conf.\ on Computer Vision and Pattern Recognition (CVPR)},
  pages     = {779--788},
  year      = {2016},
  doi       = {10.1109/CVPR.2016.91}
}

@software{ultralytics_yolo,
  author  = {Jocher, Glenn and Qiu, Jing and Chaurasia, Ayush},
  title   = {Ultralytics YOLO},
  year    = {2023},
  version = {8.0.0},
  url     = {https://github.com/ultralytics/ultralytics},
  note    = {AGPL-3.0 licensed software}
}

@article{powell2018parameter,
  title={Parameter estimation and model selection of gravitational wave signals contaminated by transient detector noise glitches},
  author={Powell, Jade},
  journal={Classical and Quantum Gravity},
  volume={35},
  number={15},
  pages={155017},
  year={2018},
  publisher={IOP Publishing}
}

@article{Samajdar2021,
  title = {Biases in parameter estimation from overlapping gravitational-wave signals in the third-generation detector era},
  author = {Anuradha Samajdar and Justin Janquart and Chris Van Den Broeck and Tim Dietrich},
  journal = {Physical Review D},
  volume = {104},
  number = {4},
  pages = {044003},
  year = {2021},
  publisher = {American Physical Society},
  doi = {10.1103/PhysRevD.104.044003},
  url = {https://doi.org/10.1103/PhysRevD.104.044003}
}

@article{Himemoto2021Impacts,
  author = {Himemoto, Yoshiaki and Nishizawa, Atsushi and Taruya, Atsushi},
  title = {Impacts of overlapping gravitational-wave signals on the parameter estimation: Toward the search for cosmological backgrounds},
  journal = {Physical Review D},
  volume = {104},
  number = {4},
  pages = {044010},
  year = {2021},
  month = {August},
  publisher = {American Physical Society},
  doi = {10.1103/PhysRevD.104.044010},
  eprint = {2103.14816},
  archivePrefix = {arXiv},
  primaryClass = {gr-qc},
  url = {https://doi.org/10.1103/PhysRevD.104.044010}
}

@inproceedings{zhou2016learning,
  author    = {Bolei Zhou and Aditya Khosla and Agata Lapedriza and Aude Oliva and Antonio Torralba},
  title     = {Learning Deep Features for Discriminative Localization},
  booktitle = {Proceedings of the IEEE Conference on Computer Vision and Pattern Recognition (CVPR)},
  pages     = {2921--2929},
  year      = {2016}
}

@article{nakamura1998gravitational,
  title={Gravitational lensing of gravitational waves from inspiraling binaries by a point mass lens},
  author={Nakamura, Takahiro T},
  journal={Physical review letters},
  volume={80},
  number={6},
  pages={1138},
  year={1998},
  publisher={APS}
}

@article{nakamura1999wave,
  title={Wave optics in gravitational lensing},
  author={Nakamura, Takahiro T and Deguchi, Shuji},
  journal={Progress of Theoretical Physics Supplement},
  volume={133},
  pages={137--153},
  year={1999},
  publisher={Oxford Academic}
}

@article{takahashi2003wave,
  title={Wave effects in the gravitational lensing of gravitational waves from chirping binaries},
  author={Takahashi, Ryuichi and Nakamura, Takashi},
  journal={The Astrophysical Journal},
  volume={595},
  number={2},
  pages={1039},
  year={2003},
  publisher={IOP Publishing}
}

@article{sathyaprakash1991,
author = {B.\~S.\~Sathyaprakash and S.\~V.\~Dhurandhar},
title = {Choice of filters for the detection of gravitational waves from coalescing binaries},
journal = {Phys. Rev. D},
volume = {44},
number = {12},
pages = {3819--3834},
year = {1991},
doi = {10.1103/PhysRevD.44.3819}
}

@article{allen2012,
author = {B.\~Allen and W.\~G.\~Anderson and P.\~R.\~Brady and D.\~A.\~Brown and J.\~D.\~E.\~Creighton},
title = {{FINDCHIRP}: An algorithm for detection of gravitational waves from inspiraling compact binaries},
journal = {Phys. Rev. D},
volume = {85},
pages = {122006},
year = {2012},
doi = {10.1103/PhysRevD.85.122006}
}

@article{usman2016_pycbc,
author = {S.\~A.\~Usman and A.\~H.\~Nitz and I.\~W.\~Harry and C.\~M.\~Biwer and D.\~A.\~Brown and others},
title = {The {PyCBC} search for gravitational waves from compact binary coalescence},
journal = {Class. Quantum Grav.},
volume = {33},
number = {21},
pages = {215004},
year = {2016},
doi = {10.1088/0264-9381/33/21/215004}
}

@article{chatterji2004_qtransform,
author = {S.\~Chatterji and L.\~Blackburn and G.\~Martin and E.\~Katsavounidis},
title = {Multiresolution techniques for the detection of gravitational-wave bursts},
journal = {Class. Quantum Grav.},
volume = {21},
number = {20},
pages = {S1809--S1818},
year = {2004},
doi = {10.1088/0264-9381/21/20/024}
}

@article{robinet2020_omicron,
author = {F.\~Robinet and N.\~Arnaud and N.\~Leroy and A.\~Lundgren and D.\~Macleod and J.\~McIver},
title = {Omicron: A tool to characterize transient noise in gravitational-wave detectors},
journal = {SoftwareX},
volume = {12},
pages = {100620},
year = {2020},
doi = {10.1016/j.softx.2020.100620}
}

@article{LIGO:2021ppb,
    author = "Davis, Derek and others",
    collaboration = "LIGO",
    title = "{LIGO detector characterization in the second and third observing runs}",
    eprint = "2101.11673",
    archivePrefix = "arXiv",
    primaryClass = "astro-ph.IM",
    reportNumber = "P2000495",
    doi = "10.1088/1361-6382/abfd85",
    journal = "Class. Quant. Grav.",
    volume = "38",
    number = "13",
    pages = "135014",
    year = "2021"
}

@article{LIGO:2024kkz,
    author = "Soni, S. and others",
    collaboration = "LIGO",
    title = "{LIGO Detector Characterization in the first half of the fourth Observing run}",
    eprint = "2409.02831",
    archivePrefix = "arXiv",
    primaryClass = "astro-ph.IM",
    doi = "10.1088/1361-6382/adc4b6",
    journal = "Class. Quant. Grav.",
    volume = "42",
    number = "8",
    pages = "085016",
    year = "2025"
}

@article{Macas:2022afm,
    author = "Macas, Ronaldas and Pooley, Joshua and Nuttall, Laura K. and Davis, Derek and Dyer, Martin J. and Lecoeuche,
              Yannick and Lyman, Joseph D. and McIver, Jess and Rink, Katherine",
    title = "{Impact of noise transients on low latency gravitational-wave event localization}",
    eprint = "2202.00344",
    archivePrefix = "arXiv",
    primaryClass = "astro-ph.HE",
    doi = "10.1103/PhysRevD.105.103021",
    journal = "Phys. Rev. D",
    volume = "105",
    number = "10",
    pages = "103021",
    year = "2022"
}

@article{Hourihane:2022doe,
    author = "Hourihane, Sophie and Chatziioannou, Katerina and Wijngaarden, Marcella and Davis, Derek and Littenberg, Tyson and Cornish, Neil",
    title = "{Accurate modeling and mitigation of overlapping signals and glitches in gravitational-wave data}",
    eprint = "2205.13580",
    archivePrefix = "arXiv",
    primaryClass = "gr-qc",
    reportNumber = "P2200160",
    doi = "10.1103/PhysRevD.106.042006",
    journal = "Phys. Rev. D",
    volume = "106",
    number = "4",
    pages = "042006",
    year = "2022"
}

@article{Glanzer_2023,
   title={Data quality up to the third observing run of advanced LIGO: Gravity Spy glitch classifications},
   volume={40},
   ISSN={1361-6382},
   url={http://dx.doi.org/10.1088/1361-6382/acb633},
   DOI={10.1088/1361-6382/acb633},
   number={6},
   journal={Classical and Quantum Gravity},
   publisher={IOP Publishing},
   author={Glanzer, J and Banagiri, S and Coughlin, S B and Soni, S and Zevin, M and Berry, C P L and Patane, O and Bahaadini, S and Rohani, N and Crowston, K and Kalogera, V and Østerlund, C and Trouille, L and Katsaggelos, A},
   year={2023},
   month=feb, pages={065004} }

@article{Soni:2021cjy,
    author = "Soni, S. and others",
    title = "{Discovering features in gravitational-wave data through detector characterization, citizen science and machine learning}",
    eprint = "2103.12104",
    archivePrefix = "arXiv",
    primaryClass = "gr-qc",
    doi = "10.1088/1361-6382/ac1ccb",
    journal = "Class. Quant. Grav.",
    volume = "38",
    number = "19",
    pages = "195016",
    year = "2021"
}

@article{LIGOScientific:2014pky,
    author = "Aasi, J. and others",
    collaboration = "LIGO Scientific",
    title = "{Advanced LIGO}",
    eprint = "1411.4547",
    archivePrefix = "arXiv",
    primaryClass = "gr-qc",
    doi = "10.1088/0264-9381/32/7/074001",
    journal = "Class. Quant. Grav.",
    volume = "32",
    pages = "074001",
    year = "2015"
}

@article{VIRGO:2014yos,
    author = "Acernese, F. and others",
    collaboration = "Virgo",
    title = "{Advanced Virgo: a second-generation interferometric gravitational wave detector}",
    eprint = "1408.3978",
    archivePrefix = "arXiv",
    primaryClass = "gr-qc",
    doi = "10.1088/0264-9381/32/2/024001",
    journal = "Class. Quant. Grav.",
    volume = "32",
    number = "2",
    pages = "024001",
    year = "2015"
}

@article{kagra,
    author = "Akutsu, T. and others",
    collaboration = "KAGRA",
    title = "{Overview of KAGRA: Detector design and construction history}",
    eprint = "2005.05574",
    archivePrefix = "arXiv",
    primaryClass = "physics.ins-det",
    doi = "10.1093/ptep/ptaa125",
    journal = "PTEP",
    volume = "2021",
    number = "5",
    pages = "05A101",
    year = "2021"
}

@article{LIGOScientific:2016aoc,
    author = "Abbott, B. P. and others",
    collaboration = "LIGO Scientific, Virgo",
    title = "{Observation of Gravitational Waves from a Binary Black Hole Merger}",
    eprint = "1602.03837",
    archivePrefix = "arXiv",
    primaryClass = "gr-qc",
    reportNumber = "LIGO-P150914",
    doi = "10.1103/PhysRevLett.116.061102",
    journal = "Phys. Rev. Lett.",
    volume = "116",
    number = "6",
    pages = "061102",
    year = "2016"
}

@article{Goode:2024ccn,
    author = "Goode, Simon R. and Schiworski, Mitchell and Brown, Daniel and Thrane, Eric and Lasky, Paul D.",
    title = "{You only thermoelastically deform once: point absorber detection in LIGO test masses with YOLO}",
    eprint = "2411.16104",
    archivePrefix = "arXiv",
    primaryClass = "astro-ph.IM",
    doi = "10.1364/OE.554989",
    journal = "Opt. Express",
    volume = "33",
    number = "8",
    pages = "17601--17616",
    year = "2025"
}

@article{LIGOScientific:2019hgc,
    author = "Abbott, Benjamin P and others",
    collaboration = "LIGO Scientific, Virgo",
    title = "{A guide to LIGO{\textendash}Virgo detector noise and extraction of transient gravitational-wave signals}",
    eprint = "1908.11170",
    archivePrefix = "arXiv",
    primaryClass = "gr-qc",
    doi = "10.1088/1361-6382/ab685e",
    journal = "Class. Quant. Grav.",
    volume = "37",
    number = "5",
    pages = "055002",
    year = "2020"
}

@article{KAGRA:2021vkt,
    author = "Abbott, R. and others",
    collaboration = "KAGRA, VIRGO, LIGO Scientific",
    title = "{GWTC-3: Compact Binary Coalescences Observed by LIGO and Virgo during the Second Part of the Third Observing Run}",
    eprint = "2111.03606",
    archivePrefix = "arXiv",
    primaryClass = "gr-qc",
    reportNumber = "LIGO-P2000318",
    doi = "10.1103/PhysRevX.13.041039",
    journal = "Phys. Rev. X",
    volume = "13",
    number = "4",
    pages = "041039",
    year = "2023"
}

@article{Aso:2013eba,
    author = "Aso, Yoichi and Michimura, Yuta and Somiya, Kentaro and Ando, Masaki and Miyakawa, Osamu and Sekiguchi, Takanori and Tatsumi, Daisuke and Yamamoto, Hiroaki",
    collaboration = "KAGRA",
    title = "{Interferometer design of the KAGRA gravitational wave detector}",
    eprint = "1306.6747",
    archivePrefix = "arXiv",
    primaryClass = "gr-qc",
    doi = "10.1103/PhysRevD.88.043007",
    journal = "Phys. Rev. D",
    volume = "88",
    number = "4",
    pages = "043007",
    year = "2013"
}

@article{aLIGO:2020wna,
    author = "Buikema, Aaron and others",
    collaboration = "aLIGO",
    title = "{Sensitivity and performance of the Advanced LIGO detectors in the third observing run}",
    eprint = "2008.01301",
    archivePrefix = "arXiv",
    primaryClass = "astro-ph.IM",
    doi = "10.1103/PhysRevD.102.062003",
    journal = "Phys. Rev. D",
    volume = "102",
    number = "6",
    pages = "062003",
    year = "2020"
}

@article{Davis:2018yrz,
    author = "Davis, D. and Massinger, T. J. and Lundgren, A. P. and Driggers, J. C. and Urban, A. L. and Nuttall, L. K.",
    title = "{Improving the Sensitivity of Advanced LIGO Using Noise Subtraction}",
    eprint = "1809.05348",
    archivePrefix = "arXiv",
    primaryClass = "astro-ph.IM",
    doi = "10.1088/1361-6382/ab01c5",
    journal = "Class. Quant. Grav.",
    volume = "36",
    number = "5",
    pages = "055011",
    year = "2019"
}

@article{Capote:2024rmo,
    author = "Capote, E. and others",
    title = "{Advanced LIGO detector performance in the fourth observing run}",
    eprint = "2411.14607",
    archivePrefix = "arXiv",
    primaryClass = "gr-qc",
    reportNumber = "LIGO-P2400256",
    doi = "10.1103/PhysRevD.111.062002",
    journal = "Phys. Rev. D",
    volume = "111",
    number = "6",
    pages = "062002",
    year = "2025"
}

@article{LIGOScientific:2016gtq,
    author = "Abbott, B. P. and others",
    collaboration = "LIGO Scientific, Virgo",
    title = "{Characterization of transient noise in Advanced LIGO relevant to gravitational wave signal GW150914}",
    eprint = "1602.03844",
    archivePrefix = "arXiv",
    primaryClass = "gr-qc",
    doi = "10.1088/0264-9381/33/13/134001",
    journal = "Class. Quant. Grav.",
    volume = "33",
    number = "13",
    pages = "134001",
    year = "2016"
}

@dataset{lvk_gwtc3_o3_sensitivity_2023,
  author       = {{LIGO Scientific Collaboration and Virgo Collaboration and KAGRA Collaboration}},
  title        = {{GWTC-3: Compact Binary Coalescences Observed by LIGO and Virgo During the Second Part of the Third Observing Run — O3 search sensitivity estimates}},
  year         = {2023},
  month        = may,
  publisher    = {Zenodo},
  version      = {v2},
  doi          = {10.5281/zenodo.7890437},
  url          = {https://doi.org/10.5281/zenodo.7890437}
}

@misc{wang2019cspnetnewbackboneenhance,
      title={CSPNet: A New Backbone that can Enhance Learning Capability of CNN}, 
      author={Chien-Yao Wang and Hong-Yuan Mark Liao and I-Hau Yeh and Yueh-Hua Wu and Ping-Yang Chen and Jun-Wei Hsieh},
      year={2019},
      eprint={1911.11929},
      archivePrefix={arXiv},
      primaryClass={cs.CV},
      url={https://arxiv.org/abs/1911.11929}, 
}

@misc{liu2018pathaggregationnetworkinstance,
      title={Path Aggregation Network for Instance Segmentation}, 
      author={Shu Liu and Lu Qi and Haifang Qin and Jianping Shi and Jiaya Jia},
      year={2018},
      eprint={1803.01534},
      archivePrefix={arXiv},
      primaryClass={cs.CV},
      url={https://arxiv.org/abs/1803.01534}, 
}

@software{gwpy_3_1_0,
  author       = {Duncan Macleod and others},
  title        = {{GWpy: A Python package for gravitational-wave astrophysics (v3.1.0)}},
  year         = 2024,
  publisher    = {Zenodo},
  version      = {3.1.0},
  doi          = {10.5281/zenodo.12734623},
  url          = {https://doi.org/10.5281/zenodo.12734623}
}

@article{Cornish:2014kda,
    author = "Cornish, Neil J. and Littenberg, Tyson B.",
    title = "{BayesWave: Bayesian Inference for Gravitational Wave Bursts and Instrument Glitches}",
    eprint = "1410.3835",
    archivePrefix = "arXiv",
    primaryClass = "gr-qc",
    doi = "10.1088/0264-9381/32/13/135012",
    journal = "Class. Quant. Grav.",
    volume = "32",
    number = "13",
    pages = "135012",
    year = "2015"
}

@article{Soni:2023kqq,
    author = "Soni, Siddharth and Glanzer, Jane and Effler, Anamaria and Frolov, Valera and Gonz{\'a}lez, Gabriela and Pele, Arnaud and Schofield, Robert",
    title = "{Modeling and reduction of high frequency scatter noise at LIGO Livingston}",
    eprint = "2311.05730",
    archivePrefix = "arXiv",
    primaryClass = "astro-ph.IM",
    doi = "10.1088/1361-6382/ad494a",
    journal = "Class. Quant. Grav.",
    volume = "41",
    number = "13",
    pages = "135015",
    year = "2024"
}

@article{Cabero:2019orq,
    author = "Cabero, Miriam and others",
    title = "{Blip glitches in Advanced LIGO data}",
    eprint = "1901.05093",
    archivePrefix = "arXiv",
    primaryClass = "physics.ins-det",
    doi = "10.1088/1361-6382/ab2e14",
    journal = "Class. Quant. Grav.",
    volume = "36",
    number = "15",
    pages = "15",
    year = "2019"
}

@ARTICLE{crop_weed_yolo,
  author={Li, Zhan and Yan, Yiming and Wei, Maosheng and Ge, Baoyu and Su, Nan},
  journal={IEEE Signal Processing Letters}, 
  title={FGOD-YOLOv8: Fine-Grained Object Detection for Crops and Weeds}, 
  year={2025},
  volume={32},
  number={},
  pages={791-795},
  doi={10.1109/LSP.2024.3519272}}

@ARTICLE{water_leak_yolo,
  author={Chen, Junxin and Xu, Xu and Jeon, Gwanggil and Camacho, David and He, Ben-Guo},
  journal={IEEE Transactions on Emerging Topics in Computational Intelligence}, 
  title={WLR-Net: An Improved YOLO-V7 With Edge Constraints and Attention Mechanism for Water Leakage Recognition in the Tunnel}, 
  year={2024},
  volume={8},
  number={4},
  pages={3105-3116},
  doi={10.1109/TETCI.2024.3369999}}

@article{yolo_obj_detection,
title = {Real-time Object Detection in Autonomous Vehicles with YOLO},
journal = {Procedia Computer Science},
volume = {246},
pages = {2792-2801},
year = {2024},
note = {28th International Conference on Knowledge Based and Intelligent information and Engineering Systems (KES 2024)},
issn = {1877-0509},
doi = {https://doi.org/10.1016/j.procs.2024.09.392},
url = {https://www.sciencedirect.com/science/article/pii/S1877050924024293},
author = {Nusaybah M. Alahdal and Felwa Abukhodair and Leila Haj Meftah and Asma Cherif}}

@dataset{glanzer2021gravity,
  author       = {Glanzer, J. and Banagari, S. and Coughlin, S. and Zevin, M. and Bahaadini, S. and Rohani, N. and Allen, S. and Berry, C. and Crowston, K. and Harandi, M. and Jackson, C. and Kalogera, V. and Katsaggelos, A. and Noroozi, V. and Osterlund, C. and Patane, O. and Smith, J. and Soni, S. and Trouille, L.},
  title        = {Gravity Spy Machine Learning Classifications of LIGO Glitches from Observing Runs O1, O2, O3a, and O3b (v1.0.0)},
  year         = {2021},
  publisher    = {Zenodo},
  doi          = {10.5281/zenodo.5649212},
  url          = {https://doi.org/10.5281/zenodo.5649212},
  type         = {dataset},
  version      = {v1.0.0}
}

@misc{LIGO2021GWTC3,
  author       = {{LIGO Scientific Collaboration and Virgo Collaboration and KAGRA Collaboration}},
  title        = {{GWTC-3: Compact Binary Coalescences Observed by LIGO and Virgo During the Second Part of the Third Observing Run — O3 search sensitivity estimates}},
  year         = {2021},
  howpublished = {[Data set] on Zenodo},
  publisher    = {Zenodo},
  doi          = {10.5281/zenodo.5546676},
  url          = {https://doi.org/10.5281/zenodo.5546676},
}

@article{Zevin:2023rmt,
    author = "Zevin, Michael and others",
    title = "{Gravity Spy: lessons learned and a path forward}",
    eprint = "2308.15530",
    archivePrefix = "arXiv",
    primaryClass = "gr-qc",
    doi = "10.1140/epjp/s13360-023-04795-4",
    journal = "Eur. Phys. J. Plus",
    volume = "139",
    number = "1",
    pages = "100",
    year = "2024"
}

@article{Wu:2024tpr,
    author = "Wu, Yunan and Zevin, Michael and Berry, Christopher P. L. and Crowston, Kevin and {\O}sterlund, Carsten and Doctor, Zoheyr and Banagiri, Sharan and Jackson, Corey B. and Kalogera, Vicky and Katsaggelos, Aggelos K.",
    title = "{Advancing glitch classification in Gravity Spy: multi-view fusion with attention-based machine learning for Advanced LIGO{\textquoteright}s fourth observing run}",
    eprint = "2401.12913",
    archivePrefix = "arXiv",
    primaryClass = "gr-qc",
    reportNumber = "LIGO DCC P2300458",
    doi = "10.1088/1361-6382/adf58b",
    journal = "Class. Quant. Grav.",
    volume = "42",
    number = "16",
    pages = "165015",
    year = "2025"
}

@software{roboflow_software_2025,
  author  = {Dwyer, Brad and Nelson, Joseph and Hansen, Thomas and others},
  title   = {Roboflow},
  version = {1.0},
  year    = {2025},
  note    = {Computer vision software},
  url     = {https://roboflow.com}
}

@misc{GW_dataset,
  author    = {Soni, Siddharth and Menon, Nikhil Mukund},
  title     = {{GW-YOLO Training Dataset}},
  year      = {2025},
  month     = sep,
  publisher = {Zenodo},
  doi       = {10.5281/zenodo.17211276},
  url       = {https://zenodo.org/records/17211276
  }
}

@unpublished{LIGO:2026ika,
  author        = {Glanzer, J. and others},
  collaboration = {LIGO},
  title         = {LIGO Detector Characterization in the Second and Third Parts of the Fourth Observing Run},
  eprint        = {2608.12193},
  archivePrefix = {arXiv},
  primaryClass  = {astro-ph.IM},
  year          = {2026}
}

@article{Benoit2025Aframe,
  author  = {Benoit, William and Marx, Ethan J. and Gunny, Alec M. and
             Omer, Rafia and Chatterjee, Deep and Saleem, Muhammed and
             Moreno, Eric and Raikman, Ryan J. and Govorkova, Ekaterina and
             Coughlin, Michael W. and Harris, Philip C. and
             Katsavounidis, Erik},
  title   = {A machine-learning pipeline for real-time detection of
             gravitational waves from compact binary coalescences},
  journal = {Physical Review D},
  volume  = {111},
  pages   = {042010},
  year    = {2025},
  doi     = {10.1103/PhysRevD.111.042010}
}

\end{document}